\documentclass[lettersize,journal]{IEEEtran}
\usepackage{xspace}
\usepackage{xcolor}
\definecolor{lightcyan}{rgb}{0.88, 1.0, 1.0}
\definecolor{pygmentsDefaultNameVariable}{HTML}{19177C}
\usepackage{graphicx}
\usepackage{subcaption}
\usepackage{microtype}
\usepackage{hyphenat}
\usepackage{booktabs}
\usepackage{multirow}
\usepackage{amsmath}
\usepackage{enumitem}
\usepackage{tikz}
\usetikzlibrary{shapes,arrows}
\usetikzlibrary{positioning, shadows}
\usetikzlibrary{fit}
\usetikzlibrary{patterns}
\usepackage{url}
\usepackage{syntax}
\usepackage{amssymb}
\usepackage{pifont}% http://ctan.org/pkg/pifont
\usepackage{pgf-pie}

\usepackage[many]{tcolorbox}

\usepackage{ebproof}

\usepackage[ruled, lined, linesnumbered, commentsnumbered, longend]{algorithm2e}

\usepackage[frozencache,cachedir=.]{minted}
\usepackage{array}
\usepackage{longtable}
\usepackage{breakurl}

\setminted[solidity]{fontsize=\footnotesize,xleftmargin=13pt,numbersep=5pt,linenos,texcomments,%
breaklines}
\setminted[prolog]{fontsize=\small,xleftmargin=13pt,texcomments,breaklines}
\AtBeginEnvironment{minted}{%
  }
\newcommand{\code}[1]{\texttt{#1}\xspace}
\newcommand{\name}[1]{\textsc{#1}\xspace}

\newcommand{\tool}{\textsc{InvCon+}\xspace}
\newcommand{\houdini}{Houdini\xspace}

\renewcommand{\paragraph}[1]{\vskip 0.05in \noindent\textbf{#1.}}

\newcommand{\OpenZeppelin}{OpenZeppelin\xspace}

\newcommand{\website}{\url{https://sites.google.com/view/invconplus/}\xspace}
\newcommand{\erc}{ERC20\xspace}

\newcommand{\eg}{e.g.\xspace}

\usepackage{listings, xcolor}

\definecolor{verylightgray}{rgb}{.97,.97,.97}

\lstdefinelanguage{Solidity}{
	keywords=[1]{anonymous, assembly, assert, balance, break, call, callcode, case, catch, class, constant, continue, constructor, contract, debugger, default, delegatecall, delete, do, else, emit, event, experimental, export, external, false, finally, for, function, gas, if, implements, import, in, indexed, instanceof, interface, internal, is, length, library, log0, log1, log2, log3, log4, memory, modifier, new, payable, pragma, private, protected, public, pure, push, require, return, returns, revert, selfdestruct, send, solidity, storage, struct, suicide, super, switch, then, this, throw, true, try, typeof, using, value, view, while, with, addmod, ecrecover, keccak256, mulmod, ripemd160, sha256, sha3}, % generic keywords including crypto operations
	keywordstyle=[1]\color{blue}\bfseries,
	keywords=[2]{address, bool, byte, bytes, bytes1, bytes2, bytes3, bytes4, bytes5, bytes6, bytes7, bytes8, bytes9, bytes10, bytes11, bytes12, bytes13, bytes14, bytes15, bytes16, bytes17, bytes18, bytes19, bytes20, bytes21, bytes22, bytes23, bytes24, bytes25, bytes26, bytes27, bytes28, bytes29, bytes30, bytes31, bytes32, enum, int, int8, int16, int24, int32, int40, int48, int56, int64, int72, int80, int88, int96, int104, int112, int120, int128, int136, int144, int152, int160, int168, int176, int184, int192, int200, int208, int216, int224, int232, int240, int248, int256, mapping, string, uint, uint8, uint16, uint24, uint32, uint40, uint48, uint56, uint64, uint72, uint80, uint88, uint96, uint104, uint112, uint120, uint128, uint136, uint144, uint152, uint160, uint168, uint176, uint184, uint192, uint200, uint208, uint216, uint224, uint232, uint240, uint248, uint256, var, void, ether, finney, szabo, wei, days, hours, minutes, seconds, weeks, years},	% types; money and time units
	keywordstyle=[2]\color{teal}\bfseries,
	keywords=[3]{block, blockhash, coinbase, difficulty, gaslimit, number, timestamp, msg, data, gas, sender, sig, value, now, tx, gasprice, origin},	% environment variables
	keywordstyle=[3]\color{violet}\bfseries,
	identifierstyle=\color{black},
	sensitive=false,
	comment=[l]{//},
	morecomment=[s]{/*}{*/},
	commentstyle=\color{gray}\ttfamily,
	stringstyle=\color{red}\ttfamily,
	morestring=[b]',
	morestring=[b]"
}

\usepackage{hyperref}
\usepackage[capitalize]{cleveref}
\crefname{section}{Sect.}{Sects.}
\Crefname{section}{Section}{Sections}
\crefname{definition}{Def.}{Defs.}
\Crefname{definition}{Definition}{Definitions}
\crefname{algorithm}{Alg.}{Algs.}
\Crefname{algorithm}{Algorithm}{Algorithms}

\makeatletter
\tikzset{
	database/.style={
		path picture={
			\draw (0, 1.5*\database@segmentheight) circle [x radius=\database@radius,y radius=\database@aspectratio*\database@radius];
			\draw (-\database@radius, 0.5*\database@segmentheight) arc [start angle=180,end angle=360,x radius=\database@radius, y radius=\database@aspectratio*\database@radius];
			\draw (-\database@radius,-0.5*\database@segmentheight) arc [start angle=180,end angle=360,x radius=\database@radius, y radius=\database@aspectratio*\database@radius];
			\draw (-\database@radius,1.5*\database@segmentheight) -- ++(0,-3*\database@segmentheight) arc [start angle=180,end angle=360,x radius=\database@radius, y radius=\database@aspectratio*\database@radius] -- ++(0,3*\database@segmentheight);
		},
		minimum width=2*\database@radius + \pgflinewidth,
		minimum height=3*\database@segmentheight + 2*\database@aspectratio*\database@radius + \pgflinewidth,
	},
	database segment height/.store in=\database@segmentheight,
	database radius/.store in=\database@radius,
	database aspect ratio/.store in=\database@aspectratio,
	database segment height=0.1cm,
	database radius=0.25cm,
	database aspect ratio=0.35,
}
\makeatother

\newcommand{\cmark}{\ding{51}\xspace}%
\newcommand{\xmark}{\ding{55}\xspace}%

\usepackage{amsthm}

\theoremstyle{definition}
\newtheorem{definition}{Definition}[section]

\usepackage{ragged2e}
\usepackage{makecell}

\newcommand{\ERCNumber}{361\xspace}
\newcommand{\NFTNumber}{10\xspace}

\tikzstyle{decision} = [diamond, draw,% fill=blue!20,
text width=2cm,  text badly centered, node distance=3cm, inner sep=0pt]
\tikzstyle{block} = [rectangle, draw,% fill=blue!20,
text width=2cm, text centered, rounded corners, minimum height=4em]
\tikzstyle{line} = [draw, -latex']
\tikzstyle{cloud} = [draw, ellipse,%fill=red!20,
node distance=2cm,
minimum height=4em, text width = 2cm, align=center]
\tikzstyle{doc}=[%
draw, minimum height=4em, minimum width=3em,
fill=white,
text width = 2cm,
text centered,
double copy shadow={shadow xshift=4pt,
	shadow yshift=4pt, fill=white, draw}
]

\newcommand{\preandpost}{pre/post-conditions\xspace}

\newcommand{\invcon}{\textsc{InvCon}\xspace}
\newcommand{\verisol}{{VeriSol}\xspace}
\newcommand{\naive}{\textit{\tool Naive}\xspace}
\newcommand{\primitive}{\textit{\tool Primitive}\xspace}
\newcommand{\wopartial}{\textit{\tool w/o Partial}\xspace}

\begin{document}

\title{Automated Invariant Generation for Solidity Smart Contracts}

\author{
	\IEEEauthorblockN{Ye Liu, Chengxuan Zhang, Yi Li} \\
	\IEEEauthorblockA{Nanyang Technological University, Singapore}\\
		\{ye.liu, chengxua001, yi\_li\}@ntu.edu.sg
}
% \orcid{0000-0001-6709-3721}
% \email{li0003ye@e.ntu.edu.sg}
% \affiliation{%
%   \institution{Nanyang Technological University}
%   \country{Singapore}
% }

% \author{Yi Li}
% \orcid{0000-0003-4562-8208}
% \email{yi\_li@ntu.edu.sg}
% \affiliation{%
%   \institution{Nanyang Technological University}
%   \country{Singapore}
% }



\maketitle

\begin{abstract}
Smart contracts are computer programs running on blockchains to automate the transaction execution between users.
The absence of contract specifications poses a real challenge to the correctness verification of
smart contracts.
Program invariants are properties that are always preserved throughout the execution, which
characterize an important aspect of the program behaviors.
In this paper, we propose a novel invariant generation framework, \tool, for Solidity smart
contracts.
\tool extends the existing invariant detector, InvCon, to automatically produce \emph{verified}
contract invariants based on both dynamic inference and static verification.
Unlike \tool, InvCon only produces \emph{likely} invariants, which have a high probability to hold,
yet are still not verified against the contract code.
Particularly, \tool is able to infer more expressive invariants that capture richer semantic
relations of contract code.
We evaluate \tool on \ERCNumber ERC20 and \NFTNumber ERC721 real-world contracts, as well as common
ERC20 vulnerability benchmarks.
The experimental results indicate that \tool efficiently produces high-quality invariant
specifications, which can be used to secure smart contracts from common vulnerabilities.
\end{abstract}

% \keywords{Smart contract, invariant detection.}
% 	<ccs2012>
% 	<concept>
% 	<concept_id>10011007.10011074.10011111.10003465</concept_id>
% 	<concept_desc>Software and its engineering~Software reverse engineering</concept_desc>
% 	<concept_significance>500</concept_significance>
% 	</concept>
% 	</ccs2012>

% \ccsdesc[500]{Software and its engineering~Software reverse engineering}
\section{Introduction}\label{sec:intro}

Smart contracts are computer programs that operate on blockchain networks.
They are used to facilitate the management of substantial financial assets and the automated
execution of agreements among multiple parties who lack inherent trust.
Notably, blockchain networks such as Ethereum~\cite{Ethereum} and BSC~\cite{BSC} are widely
recognized as leading platforms supporting smart contracts, with applications spanning
diverse domains such as supply-chain management, finance, energy, games, and digital
artworks.
%As of May 2022, the deployment of Solidity~\cite{solidity} smart contracts on Ethereum has surged to nearly 50 million, marking a 3.25x increase compared to just three years prior~\cite{Etherscan}. These smart contracts have played a pivotal role in the creation of 4,056 decentralized applications (\dapps), catering to approximately 113.86K daily active users~\cite{stateofthedapps}. However, it's crucial to acknowledge that the prevalence of defects in smart contract applications has resulted in substantial financial losses, exemplified by incidents such as the infamous DAO attack~\cite{DAOattacks}.
%SELECT count(*) FROM `bigquery-public-data.crypto_ethereum.contracts`
%WHERE DATE(block_timestamp) < "2019-05-07"
%result: 15340585
%SELECT count(*) FROM `bigquery-public-data.crypto_ethereum.contracts`
%result: 49938919
% 49938919 / 15340585 = 3.25
While smart contracts hold promise for facilitating value transfer among users, those that deviate
from their specifications may harbor bugs or vulnerabilities. Numerous implementations of ERC20
contracts diverge from common expectations, as exemplified by standard non-compliance of
\erc~\cite{chen2019tokenscope}, particularly concerning event emission, balance updates, and
the transaction fee mechanisms.

Even well-established standard ERC20 implementations exhibit inconsistencies~\cite{moon2022conformance}.
The root cause lies in the limited semantic specifications outlined in the ERC20 standard proposal
document~\cite{eip20}. Take the \code{transfer} function as an illustration---it is designed to
move a specified amount of tokens from the sender to the recipient while triggering the
\code{Transfer} event and should throw an error if the sender lacks adequate tokens for the
transfer. Nevertheless, the ERC20 proposal provides only simple textual descriptions of the
function, leading to semantic disparities across various ERC implementations and even different
versions of the same implementation.
For instance, the widely used \erc implementation from \OpenZeppelin initially did not permit a
return value for the \code{transfer} function until a later
commit,\footnote{\url{https://github.com/OpenZeppelin/openzeppelin-solidity/commit/6331dd125d8e8429480b2630f49781f3e1ed49cd}}
  causing incompatibility issues with renowned tokens like BNB, as reported by the reputable
  security company SECBIT~\cite{moon2022conformance}.
In cases where a contract necessitates
  checking the return value of an external call to a \code{transfer} function of ERC20 contracts,
  even if the \code{transfer} is successful, it may revert due to the absence of a return value,
  resulting in compatibility problems~\cite{incompatibilityissue}. However, removing the return
  value check exposes contracts to a potential vulnerability known as the \emph{fake deposit
  attack}~\cite{fakedeposit}.

Ensuring the correctness of smart contracts poses a significant challenge, especially in the
absence of contract specifications. On the one hand, the documentation for most smart contracts is
scant, with even widely recognized smart contract libraries like OpenZeppelin~\cite{OpenZeppelin,
issue3368} found to have errors and deficiencies in their documentation~\cite{Zhu2022ISS}.
On the other hand, the absence of contract specifications hampers the widespread adoption of formal
verification tools in the realm of smart contracts.
To address this issue, the commercial formal verification company
Certora\footnote{https://www.certora.com/} has adopted a crowd sourcing approach---they hosted
numerous competitions on well-known bug bounty platforms, such as
Code4Rena,\footnote{https://code4rena.com/} to engage third-party security experts in the
formulation of contract specifications.
Yet, manual creation of formal specifications for smart contracts remains costly and error-prune.

%The correctness of smart contracts is non-trivial to guarantee with the absence of contract specifications.
%On the one hand, most smart contracts have little documentation where even well-known smart contract libraries, such as OpenZeppelin~\cite{OpenZeppelin}, are found to contain errors and incompleteness in their documentation~\cite{issue3368}.
%On the other hand, the lack of contract specifications impedes the adoption of formal verification tool in smart contracts.
%To mitigate this problem, the commercial formal verification company
%Certora\footnote{https://www.certora.com/} has co-launched many competitions in well-known bug
%bounty platforms (e.g., Code4Rena\footnote{https://code4rena.com/}), with the goal to involve
%third-party security experts in writing contract specifications.

Many automated techniques~\cite{daikon,flanagan2001houdini} have been proposed to generate formal
specifications in various forms to support the testing, verification, and validation of software
programs.
Among them, program invariants, which are enduring properties maintained throughout program
execution, inherently serve as excellent candidates for enhancing and reinforcing program
specifications.
Program invariants have been used for vulnerability detection~\cite{wang2020oracle}, conformance
checking~\cite{chen2019tokenscope}, runtime protection~\cite{li2020securing}, type
checking~\cite{tan2022soltype}, and formal verification~\cite{so2020verismart,hajdu2020solc} for
smart contracts.
%\liuye{May add related works that use invariant in smart contract analysis.}
Established tools, such as Daikon~\cite{daikon}, can identify \emph{likely} program invariants for
Java programs through the execution of their test cases.
The process involves statistically inferring the invariants that hold based on predefined
templates, while discarding those refuted by the data trace records.
The complete historical transaction data of smart contracts is consistently stored on blockchains, encapsulating all execution data since contract deployment,
%. Leveraging the transparency feature, Ethereum boasts a substantial number of public contract-related transactions,
serving as a valuable data source for mining invariants.

In our prior work, \invcon~\cite{Liu2022Invcon} utilized Daikon to identify \emph{likely}
invariants for smart contracts, all of which are primitive predicates hold throughout the existing
transaction histories.
Moreover,  Liu et al.~\cite{liu2022learning} employed reinforcement learning to learn contract invariants critical to safely performing arithmetic operations, with focus on preventing integer overflow and underflow.
Despite their usefulness, the correctness of such inferred invariants remains unverified.
In particular, an invariant which holds in past transactions may not always hold in the
future---this may be due to the limited contract interactions observed in the transaction histories
so far.
%In contrast, \emph{verified} contract invariants offer established quality assurance of contract
%code.
%Smart contract projects often hire third-party security audit companies, e.g., Trail of
%Bits\footnote{\url{https://www.trailofbits.com/}} and Certora,  to review their code to exhibit
%trustworthy security and reliability assurance to the public.
%Except for vulnerability findings,
%each audit report contains formal verification results of desired contract properties, most of
%which are safety properties that can be expressed as contract invariants.

In this paper, we expand upon \invcon to generate \emph{verified} contract invariants
utilizing both dynamic inference and static verification.
We introduce a specialized invariant specification language tailored for Solidity smart contracts
and propose a novel approach for inferring high-quality verified invariants.
Specifically, we design a \houdini-like~\cite{flanagan2001houdini} algorithm to generate
verified invariants for smart contracts.
To address the explosion problem in searching for richer invariant candidates, such as implications
that prevail in ERC20 and ERC721~\cite{certora-erc20,certora-erc721,rosu-erc20-k-2023}
specifications, we introduce an iterative and incremental process for exploring these candidates
on demand.
We also apply control- and data-flow analyses to eliminate meaningless candidates and further
improve the invariant generation efficiency.
Our approach is implemented as an automated tool called \tool.
Through evaluation on \ERCNumber ERC20 contracts and \NFTNumber ERC721 real-world Solidity
contracts, we demonstrate that \tool produces comprehensive contract invariant specifications with
no false positives.
Furthermore, our analysis of real-world vulnerable ERC20 contracts underscores the potential of
\tool in safeguarding these contracts through the application of mined invariant specifications.

In summary, we make the following contributions:
\begin{itemize}
\item We introduce a comprehensive invariant specification language designed for expressing operational semantics in Solidity smart contracts. This language enables logical operations on variables of primitive types and commonly used data structures like structs, arrays, and mappings in Solidity.

\item We present a unified framework for generating \emph{verified} invariants in Solidity smart
contracts, combining dynamic invariant detection and static invariant verification. Specifically,
we develop a custom algorithm inspired by the Houdini algorithm to verify invariants for smart
contracts and introduce an iterative process to derive a richer class of
invariants.
%\yi{let's not say implication invariant which sounds too restrictive: ``to derive a richer class of
%invariants''.}

\item Our proposed approach is implemented in \tool, and its effectiveness is evaluated on
\ERCNumber ERC20 contracts and \NFTNumber ERC721 contracts, along with vulnerable ERC20 contracts
involving 25 types of vulnerabilities. The results demonstrate that \tool can generate high-quality
and comprehensive invariant specifications for smart contracts. The dataset, raw results, and the
prototype used in our experiments are available online at: \website.
\end{itemize}

\paragraph{Organizations}
The rest of the paper is organized as follows. \Cref{sec:back} provides the background about smart contracts and invariant inference.
\Cref{sec:language} defines the invariant specification language.
Then, \cref{sec:approach} introduces our invariant generation approach.
\Cref{sec:impl} describes our implementation framework, \tool, and \cref{sec:eval} demonstrates our
evaluation results.
The related work is discussed in \cref{sec:relate} and we conclude the paper in
\cref{sec:conclusion}.

\section{Background}\label{sec:back}

\subsection{Solidity Smart Contracts}

\begin{figure}[t]
	\small
	\centering
	\begin{align*}
%		Contract &::= \vec{v}; \vec{f} \nonumber\\
		a, v \in Variable &::= address \;| \; uint \;|\; int\;|\; string\; |\; bytes \;|\; \nonumber\\
		& \quad \quad  byte \;|\; bool \;|\; array\;|\; mapping\; |\; struct \{{\vec{v}} \}
		\;\nonumber\\
		f\in Function &::= func(\vec{a})\;\{{\vec{s}} \} \nonumber\\
		s \in Statement &::= v\;|\; v := e\;|\; \textbf{if}\;(e)\; \{\vec{s} \}\;\textbf{else}\;\{
		{\vec{s}} \} | \nonumber\\
				& \quad \quad \textbf{call}({\vec{e}} )\; |\; {\textbf{return}}\; e\; | \nonumber\\
		& \quad \quad\textbf{require}(e)\; |\; \textbf{assert}(e)\; |\; \textbf{revert}  \nonumber\\
		e \in Expr &::= v \;|\; \emph{const}\; |\; e[e] \;|\; e.v \;| \; e\;\bowtie \; e\nonumber
		% &\quad\quad e\;\bowtie \; e \nonumber\\
		%		boolexpr &::= \emph{true}\; |\; \emph{false}\; |\; expr\; [> < = \ge \le]\; expr\;| \nonumber\\
		%		& \quad \quad boolexpr\; and/or \;boolexpr\; |\; !\; boolexpr
		%		type &::= address \;| \; uint \;|\; int\;|\; string\; |\; bytes \;|\; byte \;|\; bool \;|\;\nonumber\\
		%		& \quad \quad type[]\;|\; mapping(type => type)\; |\nonumber\\
		%		& \quad \quad struct \{variable*\} \nonumber
	\end{align*}
	\caption{The core grammar of the Solidity language.}
	\label{fig:solidity}
\end{figure}

%\Cref{fig:solidity} shows the core grammar of the Solidity language.
%Notice that some language features, e.g., event emission, are omitted from the grammar for
%simplicity.
%%For example, \cref{fig:solidity} does not express the widely used event emission since it does not modify contract state variables, although such behaviors are important to ensure seamless integration between on-chain smart contracts and off-chain applications.
%Clearly, Solidity owns many primitive data types such as integer, string and boolean.
%Unique with other programming language like Java, Solidity does not allow floating number and has a
%special address type.
%This is because contracts directly interact with blockchain users where each user has a unique address and most contracts are developed to tokenize digital asset.
%In Solidity, a smart contract comprise a group of state variables and a set of functions.
%The statements in each function are either variable assignment, conditional statement, internal or external function call, requirement or assertion statements, reversion or return statement.
%Particularly, the ``require'' and ``assert'' statements can be used as part of the formal specifications based on which the experimental formal verification tool, SMTChecker~\cite{smtchecker} inside Solidity compiler can produce an automated mathematical proof for vulnerabilities checking, e.g., integer overflow/underflow.
%For expressions, $\bowtie$ is a binary operator containing $\{+,-, *, /, >,<,\geq,\leq, =, \ne,
%\land,\lor\}$.

\Cref{fig:solidity} presents the foundational grammar of the Solidity language, with certain
features, such as event emission, intentionally excluded for the sake of clarity.
%While \cref{fig:solidity} may not explicitly depict widely used event emission, as it does not
%modify contract state variables, it is essential to acknowledge the significance of such behaviors
%in ensuring seamless integration between on-chain smart contracts and off-chain applications.
Solidity encompasses various primitive data types, including integer, string, and Boolean.
Distinguishing itself from other programming languages like Java, Solidity does not permit
floating-point numbers and incorporates a distinctive address type.
This design choice is rooted in the interaction pattern between contracts and blockchain users,
each possessing a unique address.
Moreover, the majority of contracts are developed with the primary goal of tokenizing digital
assets.

A Solidity smart contract comprises a collection of state variables and a set of functions.
Statements within each function can take the form of variable assignments, conditional statements,
internal or external function calls, requirement or assertion statements, and reversion or return
statements. Notably, the \textit{require} and \textit{assert} statements can be employed to enforce
program invariants at runtime.
%These specifications serve as the basis for the experimental formal verification tool,
%SMTChecker~\cite{smtchecker}, within the Solidity compiler, enabling the automated generation of
%mathematical proofs for vulnerability checks, such as integer overflow/underflow.
%\yi{how is SMTChecker related to this paper?}
In the realm of expressions, $\bowtie$ denotes a binary operator encompassing $\{+,-, *, /, >,<,\geq,\leq, =, \ne, \land,\lor\}$.

\paragraph{Smart Contract Execution}
The execution of a smart contract function can be triggered by sending a blockchain transaction to
the contract address.
Typically, each transaction incorporates one or more contract calls, potentially leading to
alterations in contract state variables unless the transaction undergoes a reversion.
To ease the discussion in this paper, we model a smart contract $\mathit{SC}$ as a tuple
$(\vec{v}, \vec{f})$, where $\vec{v}$ is a vector of state
variables and $\vec{f}$ is a list of public functions.

\begin{definition}[Contract Execution]
Let $Dom(v)$ be the domain of a variable $v$ and $Dom(\vec{v}) = \underset{v \in
\vec{v}}{\prod} Dom(v)$.
Then, $\delta, \delta' \in Dom(\vec{v}) $ represent two reachable contract states.
For a function invocation $f(\vec{a})$, calling function $f$ with parameters values $\vec{a}$, we
define its high-level execution semantics as a state transition $\delta \xrightarrow[]{f(\vec{a})}
\delta'$.
%\begin{align}%	\delta \xrightarrow[]{t} \delta' &\models  post(post(\delta, {f_1}(\overrightarrow{a_1})) \mathrel{\leadsto}^* {f_k}(\overrightarrow{a_k})) \models \delta' \\
%	\delta \xrightarrow[]{f(\overrightarrow{a})} \delta' &\models (s, \delta) \mathrel{\leadsto}^* (return, \delta')
%\end{align}
%where $\delta = eval(\overrightarrow{v})$ is a reachable state of $SC$, namely the evaluated value of contract state variables $\overrightarrow{v}$ while $\overrightarrow{a}$ is the corresponding function parameters' values.
%$\models$ indicates the satisfiability relation.
%$f(\overrightarrow{a})$ is a function invocation moving contract state from $\delta$ to $\delta$' through a sequence of statement execution.
\end{definition}

Note that since a contract execution is triggered by a transaction recorded into a specific block
of the blokchain, the parameter values $\vec{a}$ also includes implicit transaction and block
parameters, e.g., \code{msg.sender} and \code{block.number}.

\paragraph{Transaction Histories}
The execution of a smart contract is intricately linked to its transaction histories on the
blockchain.
The transaction histories record every contract execution, capturing function calls, state
transitions, and modification to state variables from the contract deployment onward.
It encapsulates the evolution of the contract state, reflecting the cumulative effect of all
transactions.
%Examining the smart contract's transaction history is akin to navigating through its chronological log, providing insights into its behavior, interactions, and potential vulnerabilities.
This historical traceability is fundamental for auditing, debugging, and understanding the operational dynamics of smart contracts on the blockchain.

\subsection{Invariant Inference}
In this paper, we aim to mine \emph{contract-level} and \emph{function-level} invariant
specifications.

\begin{definition}[Function Pre/Post-conditions]
Let $f$ be a contract function, and predicates $p$ and $q$ be the pre/post-conditions of
$f$, respectively, which can be represented as a Hoare triple $\{p\}f\{q\}$.
Then the following condition should be satisfied.
\begin{align}
	\forall \delta, \forall \vec{a} \cdot \delta \models p \land  \delta \xrightarrow[]{f(\vec{a})}
	\delta' \implies \delta' \models q
\end{align}
\end{definition}

%\yi{here $p$ is the weakest precondition and $q$ is the strongest postcondition? if they are any
%  pre/postcondition, this definition may not work? please check!}
% \liuye{Here the definition does not involve whether the preconditions or postconditions are the
%weakest or strongest ones because the considered pre/postconditions are the specification
%requirements that contract function implementation must satisfy.
% In another word, the weakest preconditions or strongest postconditions of contract function
%implementation must imply the specified function pre/postconditions.
% I have added ``specified" term in this definition to avoid such confusion.
% }

\begin{definition}[Contract Invariant]
Given a smart contract \emph{SC}, its contract invariant $\mathit{I}$ is a predicate that must hold
for any contract function execution.
More formally, we have $\forall f \in SC \cdot \{I\} f \{I\}$.
\end{definition}

%\paragraph{Function Pre/Post-conditions}
%
%\paragraph{Contract Invariant}
%Because writing correct and comprehensive specifications is a quite challenging task for program developers,
%invariant inference has been proposed to automatically mine the aforementioned function pre/postconditions and contract invariants from program code and executions.

Invariant inference techniques can be broadly categorized as static and dynamic.
Static invariant inference (e.g., Houdini~\cite{flanagan2001houdini}) identifies function
pre/post-conditions and contract invariants that hold for any program execution.
On the other hand, dynamic invariant inference (e.g., Daikon~\cite{daikon}) identifies
\emph{likely} invariants that hold for specific contract executions (e.g., executions of a test
case).

Let $\Delta$ denotes a set of program executions $\{(\delta, f(\vec{a}), \delta')\}$, which bring
the contract state from $\delta$ to $\delta'$.
The \emph{likely} function \preandpost of $f$, i.e., $\{\hat{p}\}f\{\hat{q}\}$, hold for
$\Delta$ if $\forall (\delta, f(\vec{a}), \delta') \in \Delta, \delta \models \hat{p} \land \delta
\xrightarrow[]{f(\vec{a})} \delta' \implies \delta' \models \hat{q}$.
The \emph{likely} contract invariants of a smart contract is defined in a similar way, which is
omitted here for brevity.

\section{Invariant Specification Language}
\label{sec:language}

\begin{figure}[t]
	\small
	\centering
	\begin{align}
		&const \in \textbf{Int}, \textbf{Bool}, \textbf{Addr}, \textbf{Str} \;\;\; x \in \textbf{FreeVar} \;\;\; v \in \textbf{Var} \nonumber\\
		&e \in \textbf{Expr} ::= const \;|\; v\; |\; \textit{old}(v)  \; | \; len(v)\;|\; SumMap(v) \;|  \nonumber \\
		&\quad\quad\quad\quad\quad\quad \; e.x \;| \; e[x]\; |\; e\; \bowtie \; e\;\nonumber\\
		&p \in \textbf{Predicate} ::=  \bot \; |\;  e \; | \; e \implies e \nonumber \\
		& Statement ::= \textbf{Requires}\; p  \;|\; \textbf{Ensures}\; p \;| \; \textbf{ContractInv}\; p  \nonumber
		%	&p \in \textbf{Predicate} ::=  e \; | \; \forall \; x, \;   v[x] \; \bowtie \; e \;|\; e \implies e \nonumber
		%	&\quad\quad|\; \textit{SumMap}\; x_1, x_2, \cdots, x_n, \;  \sum_1^n v[x_i] \; \bowtie \; e
	\end{align}
	\caption{The invariant specification language.}
	\label{fig:speclanguage}
\end{figure}

%\paragraph{Invariant specification language}
\Cref{fig:speclanguage} introduces our invariant specification language designed for Solidity smart
contracts.
The language accommodates variables of four types: integer, Boolean, address, and string,
encompassing all primitive Solidity types illustrated in~\cref{fig:solidity}.
We facilitate two types of variables.
The first, denoted as $v$, pertains to function input parameters or contract state variables
maintained in the persistent storage of the blockchain.
The second, denoted as $x$, is reserved for free variables exclusively utilized to index structure
members or items within arrays and mappings.
Each invariant predicate is expressed as either a primitive logical expression or an implication
expression.
Furthermore, valid specification statements encompass function-level precondition invariant predicates (\textbf{Requires})  and postcondition invariant predicates (\textbf{Ensures}), and contract-level invariant predicates (\textbf{ContractInv}).

The expressions within the language may take the form of constants, variables, structure members, array items, and binary expressions.
The \textbf{old}($\cdot$) notation is employed to differentiate between the value of a variable before entering the function and its value upon exiting the function,
while \textbf{len}($\cdot$) refers to the array length or mapping size.
Additionally, the language incorporates the widely used \textbf{SumMap}($\cdot$) operator for
computing the arithmetic sum over mapping items.
The notation ``$e \bowtie e$'' represents arithmetic or logical binary operations, where the
operator ``$\bowtie$'' corresponds to the set defined in Solidity as shown in~\cref{fig:solidity}.

%where the original version of a variable is also introduced to distinguish the present version, a member of structures, item of mappings and arrays.
%$e_1\; \bowtie \; e_2$ constructs a binary expression where $\bowtie = \{+-><\geq\leq=\land\lor\}$.
%We also add the implication relation $=>$ between expressions and support widely used \textit{SumMap} operator to compute the sum over mapping items.
Utilizing this invariant language, we can articulate a diverse range of function

 and contract invariants.
To exemplify its application, we present a simple illustration.
In \cref{fig:transferFrom}, a basic \erc contract is depicted, featuring three state
variables---\code{totalSupply}, \code{balances},
\code{allows} (standing for allowances)---and a function, \code{transferFrom}.
The purpose of the \code{transferFrom} function is to transfer a specified amount of tokens from
the account addressed at \code{from} to another account at \code{to}.
An extensively studied \erc contract invariant of this example can be succinctly expressed as:
``$SumMap(balances) = totalSupply$''.
This assertion signifies that the total sum of items within the mapping variable \code{balances} must be equal to the value of \code{totalSupply}.
Additionally, the function pre/post-conditions can be articulated as follows.%
\par\noindent
{\small
\begin{align}
 &\textbf{Requires} \;\; \bot \nonumber\\
 \textcircled{1} \;& \textbf{Ensures}  \;\; to \neq 0 \implies allows[from][msg.sender] =\nonumber\\
 & \quad \quad\quad\quad old(allows[from][msg.sender]) - tokens \nonumber \\
\textcircled{2}\; & \textbf{Ensures} \;\; to \neq 0 \land from \ne to \implies balance[from] = old(balance[ \nonumber\\
 &\quad from])  - tokens\; \land\; balance[to] = old(balance[to]) + tokens\nonumber\\
\textcircled{3} \;&\textbf{Ensures} \;\;  to \neq 0 \land from = to \implies 	balance[from] =
\nonumber\\
 & \quad  \quad old(balance[from]) \land balance[to] =  old(balance[to]) \nonumber
% (4) \;& totalSuppy = old(totalSupply) \nonumber
\end{align}
}%
%\begin{table}[h]
%	\centering
%	\resizebox{.5\textwidth}{!}{
%		\begin{tabular}{p{.2cm}p{10cm}}
%			(1)& \RaggedRight{ to $\neq$ 0 $\implies$ \newline
%				allows[from][msg.sender] = old(allows[from][msg.sender]) - tokens}  \\
%			(2)& \RaggedRight{to $\neq$ 0 $\land$ from $\neq$ to $\implies$ \newline
%				balance[from] == old(balance[from]) - tokens $\land$ balance[to] == old(balance[to]) + tokens} \\
%			(3)& \RaggedRight{to $\neq$ 0 $\land$ from = to  $\implies$
%				balance[from] = old(balance[from]) $\land$ balance[to] ==  old(balance[to])} \\
%			(4)& \RaggedRight{totalSuppy = old(totalSupply)}
%		\end{tabular}
%	}
%\end{table}

In this instance, it is straightforward to ascertain that there are no preconditions for the
\code{transferFrom} function, assuming that all function preconditions are primitive predicates.
The function is characterized by three postconditions. The first postcondition \textcircled{1}
specifies that \code{allows} will undergo an update
(Line~\ref{line:allow}) when \code{to} is a non-zero address. Additionally, in cases where
\code{from} and \code{to} represent distinct addresses, the second postcondition \textcircled{2}
dictates that the balances should be adjusted accordingly (Lines~\ref{line:bfrom}–\ref{line:bto}).
Conversely, when \code{from} and \code{to} are identical, the last postcondition \textcircled{3}
emphasizes that the net effect on balance changes should be nullified.
A detailed exploration of how these invariants are mined will be provided in~\cref{sec:runningexample}.

%
%\begin{figure}[t]
%	\small
%	\centering
%	\begin{grammar}
%		<contract>  ::= var* | function*
%
%		<variable> ::= {type} \textit{id}
%
%		<function> ::= type \textit{id} "("var*")" \{statement*\}
%
%		<statement> ::= \textbf{if} ( boolexpr ) \{ statement* \} \textbf{else} \{ statement* \} |\textit{id} := expr | \textbf{call}(expr*)  | \textbf{require}(boolexpr) | \textbf{assert}(boolexpr) | \textbf{revert} | {\textbf{return}} expr
%
%		<expr> ::= \textit{id} |  \textit{const} | expr[expr] | expr.expr | expr [+-*\slash ] expr | boolexpr
%
%		<boolexpr> ::= \textit{true} | \textit{false} | expr [\textgreater \textless =  $\ge$ $\le$] expr | boolexpr and boolexpr | boolexpr or boolexpr| ! boolexpr
%
%		<type> ::= uint | int | string | bytes | bool | mapping(type =\textgreater type) | struct \{variable*\}
%	\end{grammar}
%	\caption{The core of Solidity language.}
%	\label{fig:solidity}
%\end{figure}

\begin{figure}[t]
	\small
	\begin{minted}[escapeinside=||,texcomments]{solidity}
contract ERC20 {
 // state variables
 uint totalSupply;
 mapping(address=>uint) balances;
 mapping(address=>mapping(address=>uint)) allows;
 ...
 function transferFrom(address from, address to, uint tokens) public returns (bool) {
   if (to == address(0)){
   	return false;
   }
   allows[from][msg.sender] = allows[from][msg.sender].sub(tokens); |\label{line:allow}|
   balances[from] = balances[from].sub(tokens); |\label{line:bfrom}|
   balances[to] = balances[to].add(tokens); |\label{line:bto}|
   return true;
 }
}
	\end{minted}
	\caption{A simple \erc contract.}
	\label{fig:transferFrom}
\end{figure}

%\liuye{The invariant Specification Language}
%
%We can abstract smart contracts into two aspects: interface and semantic model, without any loss of generality.
%The interface model consists of only the state variables' and functions' declaration while the semantic models specify the implementation of each function.
%Most EIP standards are actually part of interface model because they specify only the declaration of functions.
%Although these standards promote the interoperation in smart contract ecosystem, they lack strict rules to enforce semantic consistency between different contract implementations.
%
%\paragraph{Interface Model}
%
%\paragraph{Semantic Model}
%

\section{Invariant Generation Approach}
\label{sec:approach}

In this section, we present our algorithm for generating verified invariants in smart contracts and
elaborate on the techniques employed to infer implication invariants.
For simplicity in presentation, we use the term ``invariants'' to collectively denote both function \preandpost and contract invariants when explicit characterization is unnecessary.

\subsection{Algorithm}
\Cref{algo:houdini} outlines our approach to invariant generation. The algorithm takes a smart contract $\mathit{SC}$, a sequence of contract transactions $T$, and a set of invariant templates $Q$ as input. The output, denoted as $\mathit{Invs}$, comprises a set of \emph{verified} invariants, encompassing both primitive and implication invariants.

In this algorithm, $\mathit{Invs}$ is initialized as an empty set (\cref{line:inv}).
Subsequently, we initialize a set ${C}$ that encompasses all potential invariant candidates under
the given input (Line~\ref{line:q-init}), similar to Daikon's initialization process~\cite{daikon},
which instantiates all the parameterized invariant templates with concrete contract state variables
and function input variables.
For example, ``$X=Y$'' is a binary equation template where $X$ and $Y$ are placeholders that can be
filled by two concrete variables: $v_x$ and $v_y$ whenever $Dom(v_x)  \equiv Dom(v_y)$.
It is important to note that here ${C}$ excludes implication invariant candidates due to the
exponential complexity of traversing all implication candidates.
Instead, implication invariants will be generated on demand.
Moreover, the execution trace set ${\Delta}$ is initialized as an empty set
(\cref{line:d-init}).

The algorithm processes the transaction histories to extract corresponding execution traces.
For each transaction $t_i$, the algorithm parses it to extract the invoked function $f$ and
parameters values $\vec{a}$ (Line~\ref{line:parse}).
Additionally, the old and present contract states (i.e., values of the contract state variables),
denoted as $\delta$ and $\delta'$, respectively, are recorded.
The tuple ($\delta$, $f(\vec{a})$, $\delta'$) is added to the execution trace set $\Delta$
(\cref{line:data-add}).

Next, the algorithm executes the dynamic invariant detection procedure \name{InvDetect}  (\cref{line:inv-detect}) to obtain two classes of invariant candidates:
\begin{itemize}
	\item $C_{likely}$, likely invariant candidates that hold for the entire transaction histories.
	\item $C_{partial}$, partially supported invariant candidates that hold for a subset of transaction histories.
\end{itemize}
Subsequently, a primitive invariant inference technique, detailed in~\cref{sec:invinfer}, is
applied to infer the standing invariants out of $C_{likely}$, and all the verified invariants are
included in $\mathit{Invs}$ (\cref{line:inv-verify}).
The unverified likely invariant candidates, $C_{likely} \setminus \mathit{Invs}$, and
$C_{partial}$ are used to derive implication candidates assigned to $C_{imp}$
(\cref{line:implication}) via \name{FindImplications}, which will be detailed in
\cref{sec:implication}.
Additionally, it is important to note that the found implications may not always hold.
An iterative process is in place to validate these implications (Line~\ref{line:inv-verify2}) or
weaken these implications via \name{WeakenImplications} (Line~\ref{line:implication-2}) to identify
new ones.
This iterative process continues until all valid candidates are examined
(Line~\ref{line:loop2-start}). Finally, the algorithm returns $\mathit{Invs}$, which includes all
the correctly mined invariants from transaction histories (\cref{line:return}).

\begin{algorithm}[t]
\small
\caption{Contract Invariant Inference}\label{algo:houdini}
\SetKwInOut{KwIn}{Inputs}
\SetKwInOut{KwIgnore}{}
\SetKwInOut{KwOut}{Outputs}
\SetKwComment{Comment}{//}{}
\KwIn{$SC = \{\vec{v},\vec{f}\}$, where  each element $v_i \in \vec{v}$  is a contract state
variable and each element $f_i \in \vec{f}$ is a public contract function;}
\KwIgnore{$T=\{t_i|1 \le i \le n\}$, where each element $t_i$ is a contract transaction;}
%\KwIgnore{$F = \{f: (p_1, \cdots, p_k) \}$, where each element $f$ is a public mutable function having $k$ parameters;
%}
\KwIgnore{$Q$, a set of invariant templates.
}
\KwOut{$\mathit{Invs}$, a set of verified invariants.}
$ Invs := \emptyset$\; \label{line:inv}
$ C := \name{InitializeCandidates}(\vec{v}, \vec{f}, Q)$ \Comment*[l]{primitive candidates} \label{line:q-init}
$ \Delta := \emptyset$ \Comment*[l]{execution trace set}  \label{line:d-init}
%$ i := 1 $\; \label{line:i}
\ForEach{ $t_i \in T$ }{ \label{line:loop-start}
	$(\delta,  f(\vec{a}), \delta') \gets \name{Parse}(t_i)$ \; \label{line:parse}
%	\Comment*[r]{$f$: the called function; $\vec{a}$: the parameters values, $\delta$: the value of
%original state variables \{$\textbf{old}(v_i)| v_i \in \vec{v} $\}, and $\delta'$: the value of
%present state variables $\overrightarrow{v}$.}
	$\Delta \gets \Delta \cup (\delta, f(\overrightarrow{a}),  \delta')$\; \label{line:data-add}
} \label{line:loop-end}
	$C_{likely}$,  $C_{partial}$ $\gets$\name{InvDetect}($\Delta$, $C$)\; \label{line:inv-detect}
%	\Comment*[r]{$\{p |p \in Q, \forall d \in  D, p\downharpoonright d \equiv true \}$ for $C_{likely}$, and $\{p |p \in Q, \exists d \in  D, p\downharpoonright d \equiv true \}$ for $C_{partial}$.}  \label{line:inv-detect}
%$Q \gets Q \setminus LI, FI$ \; \label{line:q-reduce}
%\If{$LI \neq \emptyset$}{
	$\mathit{Invs} \gets \name{StaticInfer}(C_{likely})$ \; \label{line:inv-verify}
%}
$C_{imp} \gets \name{findImplications}(C_{likely} \setminus \mathit{Invs}, C_{partial}) $
\Comment*[l]{implication candidates}  \label{line:implication}
\While{$C_{imp} \ne \emptyset $}{ \label{line:loop2-start}
$\mathit{Invs} \gets \mathit{Invs} \cup \name{StaticInfer}(C_{imp})$ \; \label{line:inv-verify2}
$C_{imp} \gets \name{weakenImplications}(C_{imp} \setminus \mathit{Invs}) $\;
\label{line:implication-2}
} \label{line:loop2-end}
\Return{ $\mathit{Invs}$ } \label{line:return}
\end{algorithm}

%\begin{algorithm}[t]
%	\caption{\name{findImplications}}\label{algo:implication}
%	\small
%	\SetKwInOut{KwIn}{Input}
%	\SetKwInOut{KwIgnore}{}
%	\SetKwInOut{KwOut}{Output}
%	\SetKwComment{Comment}{//}{}
%	\KwIn{$UnprovedLikelyInvs$, a set of unproved likely invariant candidates.}
%	\KwIgnore{$PartialInvs$, a set of partially supported invariant candidates.}
%	\KwOut{${Implications}$, a set of implication invariant candidates.}
%	$Implications := \{ (\eta \implies \tau) \; | \eta \in  UnprovedLikelyInvs \cup PartialInvs, \tau \in UnprovedLikelyInvs \cup PartialInvs, \eta \neq \tau\}$ \label{line:p}\;
%	\ForEach{ $(\eta \implies \tau) \;\;\in Implications $}{
%		$V_1 \gets vars(\eta)$\;
%		$V_2 \gets vars(\tau)$\;
%		\If{$\nexists\; a \in V_1\; b \in V_2,$ b is control/dataflow-dependent on a}{
%			$Implications \gets Implications \setminus (\eta \implies \tau)$
%		}
%	}
%%	perform dependency analysis to refine $P$ by ruling out infeasible implications\;
%	\Return{$Implications$}
%\end{algorithm}

\begin{figure}[t]
	\centering
  \small
%	\resizebox{\columnwidth}{!}{%
	\begin{prooftree}
		\Hypo{\bot}
		\infer1[Init] {C_{imp} := \{ (\eta \implies \tau) \; | \eta, \tau \in  {C_{likely} \setminus
		Invs} \; \cup \; C_{partial}, \eta \neq \tau\}}
		\end{prooftree}
%	}
		\vfil
		\vspace{1em}
%		\resizebox{\columnwidth}{!}{%
		\begin{prooftree}
		\Hypo{(\eta \implies \tau) \in C_{imp}}
		%	\Hypo {(\eta_2 \implies \tau) \in FalImpls}
		\Hypo {\begin{matrix}
				\forall a \in vars(\eta) ,  \forall b \in vars(\tau). \\
				 \lnot\; dep(a, b)
				\end{matrix}
				}
		%	\Infer [rule style=no rule] 1 {R}
		%	\Hypo {[Q]}
		%	\Infer [rule style=no rule] 1 {R}
		\Infer2[Delete]{C_{imp} \gets C_{imp} \setminus (\eta \implies \tau)}
%		\Infer2{\emph{return}\;\; Cands}
	\end{prooftree}
%		}
	\caption{\name{FindImplications}}
	\label{fig:find}
\end{figure}

%\paragraph{How to choose implications}
\begin{figure}[t]
	\centering
  \small

	\begin{prooftree}
		\Hypo{\bot}
		\Infer1 [Init]{\hat{C_{imp}} := \emptyset}
	\end{prooftree}
	\vfil
	\vspace{1em}
	\resizebox{\columnwidth}{!}{%
	\begin{prooftree}
		\Hypo{(\eta_1 \implies \tau), (\eta_2 \implies \tau) \in { C_{imp} \setminus Invs}}
		%	\Hypo {(\eta_2 \implies \tau) \in FalImpls}
		\Hypo {\eta_1\land \eta_2 \not \equiv false}
		%	\Infer [rule style=no rule] 1 {R}
		%	\Hypo {[Q]}
		%	\Infer [rule style=no rule] 1 {R}
		\Infer2[Append-1]{\hat{C_{imp}} \gets \hat{C_{imp}} \cup (\eta_1 \land \eta_2 \implies \tau)}
	\end{prooftree}
	}
	\vfil
	\vspace{1em}
	\resizebox{\columnwidth}{!}{%
	\begin{prooftree}
		\Hypo{(\eta \implies \tau_1), (\eta \implies \tau_2) \in {C_{imp} \setminus Invs}}
		%	\Hypo {(\eta_2 \implies \tau) \in FalImpls}
		\Hypo {\tau_1 \lor \tau_2 \not \equiv true}
		%	\Infer [rule style=no rule] 1 {R}
		%	\Hypo {[Q]}
		%	\Infer [rule style=no rule] 1 {R}
		\Infer2[Append-2]{\hat{C_{imp}} \gets \hat{C_{imp}} \cup (\eta \implies \tau_1 \lor \tau_2)}
%		\Infer3{\emph{return}\;\; Cands}
	\end{prooftree}
	}
	\caption{\name{WeakenImplications}.}
	\label{fig:weaken}
\end{figure}

\subsection{Primitive Invariant Inference}\label{sec:invinfer}

\Cref{algo:staticinfer} illustrates our \houdini-like algorithm to infer verified primitive
invariants from the candidates mined from contract transaction histories.
First, we enable all the candidates in $SC$ via contract instrumentation
(Line~\ref{line:instrument}); each candidate is explicitly labeled by the added keywords, e.g., \textbf{ContractInv} for contract invariant,  \textbf{Requires} for function precondition, and \textbf{Ensures} for function postcondition.
Next, we invoke a modular verifier to statically verify these enabled candidates
(Line~\ref{line:verify}), i.e., verifying each function in isolation where all the corresponding
candidates are examined against the function implementation.
%\yi{each function is verified in isolation where the postconditions
%are checked assuming the preconditions and contract invariants given.}
%\liuye{the claim ``where the postconditions
%	are checked assuming the preconditions and contract invariants given" does not hold. In my
%understanding of Houdini, it will generate a verification condition for a function and then check
%it against both the annotated pre/post and contract invariants to find possible violations.}
When there is a failed invariant candidate $c$ violating the verification condition, $c$ will be
disabled in $SC$ (Line~\ref{line:disable}).
This process will continue until all the enabled candidates are verified successfully
(Line~\ref{line:correct-1}) and then returned (Line~\ref{line:correct-2}).
Particularly, whenever there is a failed assertion in $SC$, i.e., a violated condition $e$ in the
\textbf{assert}($e$) statement, %\yi{where do assertions come from?
%never mentioned before}
the algorithm terminates with an error raised (Line~\ref{line:raise}).
This happens in Solidity contracts, because \textbf{assert}($e$) is often misused to replace
\textbf{require}($e$) that enforces program requirements due to their similar effects on
transaction reversion.
%\yi{what do these have to do with your candidates? you never explain how assert or require are used
%in the instrumentation!}
For smart contracts without failed assertions, the verified invariants is a maximal subset of the
candidates whose conjunction is an inductive invariant.

\begin{algorithm}[t]
	\small
	\caption{\textsc{StaticInfer}(Candidates)}\label{algo:staticinfer}
%	\SetKwInOut{KwIn}{Inputs}
%	\SetKwInOut{KwIgnore}{}
%	\SetKwInOut{KwOut}{Outputs}
	\SetKwComment{Comment}{//}{}
%	\KwIn{$SC = \{\vec{v},\vec{f}\}$, where  each element $v_i \in \vec{v}$  is a contract state
%		variable and each element $f_i \in \vec{f}$ is a public contract function;}
%	\KwIgnore{$C$, invariant candidate set;}
	%\KwIgnore{$F = \{f: (p_1, \cdots, p_k) \}$, where each element $f$ is a public mutable function having $k$ parameters;
		%}
%	\KwOut{$C_{verified}$, a set of verified invariant.}
	Instrument $SC$ to \textbf{enable} each candidate from Candidates\; \label{line:instrument}
	\While{true}{
		result = \textsc{ModularVerify}($SC$) \; \label{line:verify}
	    \uIf{result = CORRECT}{ \label{line:correct-1}
	    	\textbf{return} enabled candidates   \Comment*[l]{verified invariants}  \label{line:correct-2}
	    }
	    \uElseIf{result = INCORRECT due to failed candidate c}{
	    	\textbf{disable} $c$ in $SC$\; \label{line:disable}
	    }
	    \Else{
	     	\textbf{raise} Error  \Comment*[l]{INCORRECT due to failed assertion in $SC$}
	     	\label{line:raise}
	    }
	}

\end{algorithm}

\subsection{Implication Invariant Inference}
\label{sec:implication}

\Cref{fig:find,fig:weaken} illustrate the two procedures for identifying implication
candidates, respectively.
In \cref{fig:find}, \name{FindImplications} employs two straightforward inference rules.
The first rule explores all the potential implication candidates from the unverified likely
invariants ${C_{likely}\setminus Invs}$ and partial invariant candidates $C_{partial}$, including
them in $C_{imp}$.
An implication invariant takes the form of $\eta \implies \tau$, where $\eta$ and $\tau$ comes from
the existing the unverified and partial invariant candidates.

However, not all of the implication candidates constructed this way are relevant in terms of
the contract semantics.
%However, it is essential to note that although $IC$ contains complete  implication predicates, not all of these implications are relevant in terms of program
%semantics.
%\yi{what does complete mean?}
%\liuye{``complete'' says $IC$ includes every possible implication candidates derived by the combination between ${C_{likely}^{refuted}}$ and $C_{partial}$.
%	Although, the term used is not correct since $IC$ here includes only simple implication predicates whose pre- or post-conditions are only single primitive predicate.
%	This transition sentence aims to highlight the necessity of eliminating meaningless implications.
%	Should we remove this sentence?
%}
An implication can possibly hold (i.e., relevant) if its precondition and postcondition align with
the data/control-flow of the contracts, and irrelevant implications should be discarded.
The notation $vars(p)$ represents variables appearing in an invariant predicate $p$; for instance,
$vars(p) = \{from, to\}$ when $p$ is ``$from \neq to$''.
Additionally, $\textbf{dep}(a, b)$ denotes whether variable $a$ depends on
variable $b$ in terms of control-flow or data-flow in smart contract functions.
To determine the valid implications, we leverage the well-known static analysis tool Slither~\cite{slither} to trace data-flow and control-flow in smart contract functions. Therefore, in~\cref{fig:find}, a \emph{Delete} rule is applied to eliminate implications that do not adhere to the data-flow and control-flow relationship. This rule is iteratively applied until no further implications can be eliminated.

Some implication candidates may be too strong and cannot be proved.
\Cref{fig:weaken} illustrates how we derive a weaker set of implication candidates $\hat{C_{imp}}$
from those unverified implication candidates denoted as ${C_{imp} \setminus Invs}$.
In \cref{fig:weaken}, \name{WeakenImplications} comprises three inference rules.
It initially sets $\hat{C_{imp}}$ to an empty set.
Then the rules \emph{Append-1} and \emph{Append-2} generate weaker implications by combing two
unverified implication candidates.
In essence, $\eta_1 \land
\eta_2 \implies \tau$ is weaker than either $\eta_1 \implies \tau$ or $\eta_2 \implies \tau$.
Similarly, $\eta \implies \tau_1 \lor \tau_2$ is weaker than both $\eta \implies \tau_1$ and
$\eta \implies \tau_2$.
To eliminate useless implications that are tautologies, we impose restrictions on the
original implications, such as $\eta_1 \land \eta_2 \not \equiv \text{false}$ and $\tau_1 \lor
\tau_2 \not \equiv \text{true}$.
It is evident that the weaker implications are also relevant as they satisfy the same
control/data-flow dependencies as the original ones.

\subsection{Termination}
The termination of \cref{algo:houdini} can be ensured by the fact that \tool can only produce a finite set of primitive invariant predicates. The conclusion regarding the termination of \cref{algo:houdini} hinges on whether the loop (Lines~\ref{line:loop2-start}-\ref{line:loop2-end}) comes to an end. In each iteration of the loop, we possess at least one implication candidate, constructed by \name{WeakenImplications} (refer to \cref{sec:implication}).
Regarding \name{WeakenImplications}, it consistently generates weaker implication candidates than
the previous ones, utilizing conjunctions over premises or disjunctions over consequences.
Assuming \name{InvDetect} yields $n$ primitive invariant predicates $C_{likely} \cup C_{partial} =
\{p_1, \ldots, p_n\}$, then the weakest implication will be at least as strong as $p_1 \land \cdots
\land p_n \implies p_1 \lor \cdots \lor p_n$. Consequently, the loop will finish in no more than $2
\times n$ iterations, establishing the termination of this algorithm.

\subsection{Running Example}
\label{sec:runningexample}

We illustrate our algorithm using the example presented in~\cref{fig:transferFrom}.
The details regarding our transaction parsing and invariant detection will be elaborated
in~\cref{sec:impl}.
For the sake of simplicity in the illustration, assume that we have already acquired a set of
likely and partially supported invariants through invariant detection on the transaction histories.
In \cref{tab:example}, the invariants labeled with \cmark are successfully verified by the static
verifier, while the ones with \xmark are unverified.
%\yi{how about \xmark?}
In Step \textcircled{1}, we perform a \houdini-like static inference on these detected invariant
candidates.
Consequently, three likely invariants are verified, excluding ${to \neq 0}$.
In the subsequent step (Step \textcircled{2}), nine additional implication invariant candidates are
generated from the previously unverified likely invariants and partially supported invariants,
according to the rules in \name{FindImplications} (see \cref{fig:find}).
However, after the modular verification, only one implication is confirmed.
Furthermore, we weaken these unverified implication invariants in Step \textcircled{3} using
\name{WeakenImplications} (see \cref{fig:weaken}) to derive four new implication candidates for
further validation.
Eventually, all the invariants listed in~\cref{sec:language} are successfully recovered (in
a logically equivalent form).
Moreover, two other invariants, $balances[to] \ge old(balances[to])$ and $balances[from]
\le old(balances[from])$, are verified, which provide additional insights on how the balances of
the sender and the receiver should change when \code{transferFrom} is called, beyond the
standard specifications.

%We illustrate our algorithm using the example shown in~\cref{fig:transferFrom}.
%The detail about our transaction parsing and invariant detection will be presented in~\cref{sec:impl}.
%To simplify the illustration,
%suppose we have obtained a set of likely invariant and partially supported invariants through invariant detection on its transaction histories.
%In \cref{tab:example}, the invariants labeled \cmark are proved by our static checker.
%In Step 1, we perform static verification on these invariants.
%As a result three likely invariants are proved except for \emph{to $\neq$ 0}.
%Next, seven more implication invariant candidates are generated in Step 2 out of the previous unproved likely invariants and partially supported invariants.
%However, only one implication is correct after validation.
%Further, we can weaken those unproved implication invariants to obtain new implication candidates.
%Finally, all the invariants listed in~\cref{sec:language} are successfully recovered.

%Not all likely invariants are verified true, \eg, \emph{to $\neq$0} is not true.
%More interestingly, we can find only one provable invariant in the initial implication invariant candidates but four more provable invariants in its weaker version.
%These provable invariants cover all the invariants listed in~\cref{sec:language}.
%Moreover, two additional invariants \emph{balances[to] $\ge$ old(balances[to])} and \emph{balances[from] $\le$ balances[from]} demonstrate the monotonic feature, which may generalize to every different implementation about ``transferFrom'' function.

\begin{table*}[t]
	\centering
	\caption{Illustration example of invariant verification.}
	\small
	\scalebox{.9}{
	\begin{tabular}{p{.05\textwidth} p{.35\textwidth}p{.54\textwidth}}
		\toprule
		\textbf{Step} & \textbf{Invariants} & \\
		\midrule
		\textcircled{1}& \RaggedRight{
			\textbf{Likely Contract Invariants}: \newline
		totalSupply = SumMap(balances) \cmark \newline
			\textbf{Likely Function Pre/post-conditions}: \newline
%		totalSupply = old(totalSupply) \cmark \newline
		to $\neq$0 \xmark \newline
		balances[to] $\ge$ old(balances[to]) \cmark \newline
		balances[from] $\le$ old(balances[from]) \cmark
		} &
		\RaggedRight{
			\textbf{Partially Supported Function Pre/post-conditions}: \newline
			from $\neq$ to \newline
			from = to \newline
   		    balances[from] = old(balances[from]) - tokens \newline
   		    balances[to] = old(balances[to]) + tokens \newline
   		    allows[from][msg.sender] = old(allows[from][msg.sender]) - tokens \newline
   		    balances[from] = old(balances[from]) \newline
   		    balances[to] = old(balances[to])
		}
		\\
%			\midrule
%		& Likely Invariants & Falsified  Invariants (but with minimum support)  \\
%		&totalSupply = SumMap(old(balances)) \cmark  & 	from $\neq$to    \\
%		& totalSupply = old(totalSupply) \cmark &  from = to  \\
%	 	& to $\neq$0 &  balances[from] = old(balances[from]) - tokens \\
%	    & balances[to] $\ge$ old(balances[to])  \cmark &balances[to] = old(balances[to]) + tokens  \\
%%		 & balances[from] = old(balances[from])   \\
%%		 & balances[to] = old(balances[to]) \\
%		& balances[from] $\le$ old(balances[from]) \cmark & allows[from][msg.sender] = old(allows[from][msg.sender] ) - tokens \\
%		\bottomrule
		\midrule
		\textcircled{2} &
		\multicolumn{2}{p{.89\textwidth}}{
				\RaggedRight{
						\textbf{Implication Invariant Candidates:} \newline
						to $\neq$0 $\implies$ allows[from][msg.sender] = old(allows[from][msg.sender]) - tokens \cmark \newline
						to $\neq$0 $\implies$ balances[from] = old(balances[from]) - tokens \xmark \newline
						to $\neq$0 $\implies$ balances[to] = old(balances[to]) + tokens \xmark \newline
						to $\neq$0 $\implies$ balances[from] = old(balances[from] \xmark \newline
						to $\neq$0 $\implies$ balances[to] = old(balances[to]) \xmark \newline
						from $\neq$to $\implies$ balances[from] = old(balances[from]) - tokens \xmark \newline
						from $\neq$to $\implies$ balances[to] = old(balances[to]) + tokens \xmark \newline
						from = to $\implies$ balances[from] = old(balances[from] \xmark \newline
						from = to $\implies$ balances[to] = old(balances[to]) \xmark
				}
		}
		\\
		\midrule
		\textcircled{3} &
		\multicolumn{2}{p{.89\textwidth}}{
			\RaggedRight{
				\textbf{Weakened Implication Invariant Candidates:} \newline
				to $\neq$0 $\land$ from $\neq$to $\implies$ balances[from] = old(balances[from]) - tokens \cmark \newline
				to $\neq$0 $\land$ from $\neq$to $\implies$ balances[to] = old(balances[to]) + tokens \cmark \newline
				to $\neq$0 $\land$ from = to $\implies$ balances[from] = old(balances[from]) \cmark \newline
				to $\neq$0 $\land$ from = to $\implies$ balances[to] = old(balances[to]) \cmark
			}
		}
		\\
		\bottomrule
	\end{tabular}
}
\label{tab:example}
\end{table*}

\section{Implementation}\label{sec:impl}

\subsection{Overview}
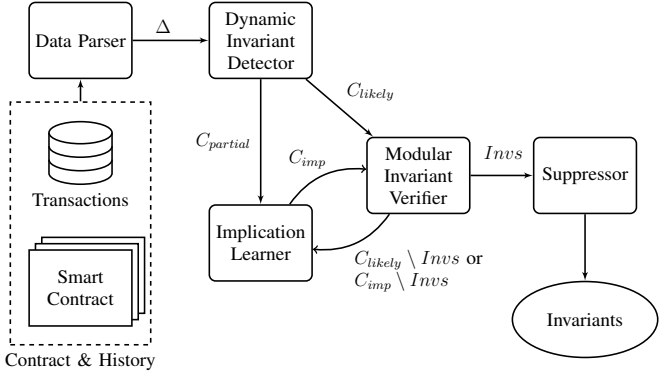
\begin{figure}[t]
\large\centering
\resizebox{\columnwidth}{!}{%
\begin{tikzpicture}[node distance = 2cm, line width=1pt, auto, fill fraction/.style={path picture={
			\fill[#1]
			(path picture bounding box.south) rectangle
			(path picture bounding box.north west);
	}},
	fill fraction/.default=gray!50]
	% Place nodes
	\node[database, label=below:Transactions,database radius=.7cm,database segment height=0.3cm]
	(init)  {};
	\node[doc, below of=init, node distance=3cm] (contract) {Smart Contract};
  \node[above of= init, node distance=0.5cm] (extractor) {};
  \node[block, above of= extractor, node distance = 2cm] (preprocessing) {Data
  Parser};

%    \node[cloud, right of= extractor, node distance = 3cm, yshift=.9cm] (output) {Transaction Output};
%    \node[cloud, right of= extractor, node distance = 3cm, yshift=-.9cm] (input) {Transaction Input};
%    \node[ below of= extractor, node distance = 3cm] (compiler) {};
%    \node[cloud, right of= compiler, node distance = 3cm, yshift=.6cm] (abi) {Contract ABI};
%    \node[cloud, right of= compiler, node distance = 3cm, yshift=-.9cm] (layout) {Storage Layout};
%    \node[ right of= input, yshift=-1cm, node distance = 0cm] (replayer) {};
%%    \node[cloud, right of= replayer, node distance = 3cm] (traces) {VM Traces};
%
	 \node[draw, dashed, inner sep=4mm,label=below:Contract \& History,fit=(init) (contract) (init)
	 (contract)] (history) {};

%	  \node[draw, dashed, inner sep=4mm,label=above:Data Preprocessing,fit=(output) (layout) (output) (layout)] (preprocessing) {};

%    \node[cloud, right of= preprocessing, node distance = 4.5cm] (triple) {Triples $\{(\gamma, \delta, \gamma')\}$};
    \node[block, right of= preprocessing, node distance = 4cm] (detector) {Dynamic
    Invariant Detector};
%     \node[doc, above of= detector,node distance = 2.5cm] (templates) {Invariant Templates};
%    \node[cloud, right of= detector, node distance = 3cm, fill fraction] (candidates) {Falsified/Likely};
    \node[block, right of= detector, node distance = 3.5cm, yshift=-3cm] (verifier) {Modular
    Invariant Verifier};
    \node[block, right of= verifier, node distance = 3.7cm] (suppressor) {Suppressor};
    \node[cloud, below of= suppressor, node distance = 3.2cm] (annotation) {Invariants};
%     \node[cloud, below of= verifier, node distance = 2.5cm] (result) {Disapproved Invariants};
     \node[block, below of= detector,node distance = 4.5cm] (teacher) {Implication Learner};
%     \node[cloud, left of= teacher, node distance = 3cm] (implication) {Implication Candidates};

%      \node[draw, dashed, inner sep=4mm,label=above:Invariant Detection,fit=(templates) (detector) (candidates) (detector)] (detection) {};
%      \node[draw, dashed, inner sep=4mm,label=below:Implication Learning,fit=(teacher) (implication) (teacher) (implication)] (learning) {};
%
%       \node[draw, dashed, inner sep=1.5mm,label=below:Invariant Verification,fit=(verifier) (result) (verifier) (result)] (verification) {};

%        \node[draw, ellipse,%fill=red!20,
%        node distance=2cm,
%        minimum height=2em, text width = 1cm, align=center,  above of= verifier, node distance = 3cm, xshift=1cm, fill fraction] (legend) {};
%        \node [right of=legend, node distance = 3cm] (label) {Falsified/Likely Invariants};
%        \node[draw, dashed, inner sep=1.5mm,label=above:Legend,fit=(legend) (label) (legend) (label)] {};

    \path [line] (history) -- (preprocessing);
%     \path [line] (preprocessing) -- (triple);
     \path [line] (preprocessing) -- node [text width=1cm, xshift=0.2cm, yshift=0.0cm] {$\Delta$} (detector);
%     \path [line] (templates) -- (detector);
     \path [line] (detector) -- node[text width=1.7cm] {$C_{likely}$} (verifier);
      \path [line] (detector) -- node[text width=1cm, xshift=-1.6cm] {$C_{partial}$} (teacher);
      \path [->] (verifier) edge[bend left]  node [xshift=-.2cm, text width=3cm] {$C_{likely} \setminus Invs$ or $C_{imp} \setminus Invs$ } (teacher);
      \path [->] (teacher) edge[bend left]  node[text width=1cm, xshift=.4cm, yshift=.0cm]  {$C_{imp}$} (verifier);

%     \path [line] (candidates) -- (verifier);
%     \path [line] (verifier) -- node [near start] {yes} (annotation);
     \path [line] (verifier) -- node [text width=1cm, xshift=0.1cm, yshift=0.2cm] {$Invs$}
     (suppressor);
     \path [line] (suppressor) -- (annotation);
%     \path [line] (result) -- (teacher);
%     \path [line] (teacher) -- (implication);
%     \path [line] (implication) -- (detector);
%     \path [line] (candidates) -- (teacher);
\end{tikzpicture}
}
\caption{The architecture overview of \tool.}
\label{fig:arc}
\end{figure}

\Cref{fig:arc} demonstrates the high-level architecture of \tool, our automated invariant detection
tool for Solidity smart contracts.
The inputs to \tool include a set of historical transactions and the corresponding contract source
code, while its output is a collection of smart contract invariant specifications or the
accordingly annotated contract code.
\tool comprises four modules: (1) a \emph{data parser} that decodes contract code and transaction
histories to extract concrete execution trace set; (2) a \emph{dynamic invariant detector} that
generates a set of likely and partially supported invariants; (3) a \emph{modular invariant verifier} and an
\emph{implication learner} that verify and learn contract invariants, respectively; and (4) a
\emph{suppressor} that simplifies the results by removing redundant invariants.
Notably, the implication learner has already been detailed in~\cref{sec:implication}.

\subsection{Data Parser}
Given a contract, we first collect all of its historical transactions.
For each transaction, we decode the specific function input based on the contract's Application Binary Interface (ABI), and we interpret the transaction output in accordance with the contract's storage layout specifications.
This layout dictates where each state variable is stored in the blockchain database. For instance,
as shown in~\cref{fig:solidity}, the first declared state variable \code{totalSupply} is stored at
the first slot (\code{0x0}) in the contract's blockchain database.

The input of a contract transaction is represented as a tuple \emph{({sender}, {function}, {parameters})}, which encapsulates the transaction's sender, the invoked function's name, and the corresponding input parameters.
Conversely, the transaction's output is denoted as \emph{({status}, {storageChanges})}.
Here, \emph{status} signifies the transaction's success or failure, while \emph{storageChanges} details the alterations in the contract's storage across various slots. By aligning storage slots with the contract’s storage layout, one can effectively interpret these storage modifications as changes in the values of the contract's state variables.
%, \liuye{without any precision loss.
Employing the previously described preprocessing technique enables the extraction of a sequence of data triples (i.e., execution traces). These triples consist of the actual values of state variables and function input variables at the point of function entry, as well as the most recent values of state variables at the point of function exit.
It is important to note that any misrecognition of variables can lead to incorrect invariant results.
We have implemented measures to ensure the accuracy of variable recognition. For state variables of primitive types, we directly ascertain their values, as the storage layout for these variables remains constant during runtime.
In the case of non-primitive, dynamic state variables, to reduce computational cost, we initially utilize the known variable values to hypothesize a correlation between the altered storage slots and the dynamic state variables.
However, if this approach fails to produce an accurate mapping, it becomes necessary to replay the entire transaction. This replay process enables us to track the comprehensive execution information, including storage modifications, thus allowing for the accurate determination of the correct mapping.

\subsection{Dynamic Invariant Detector}
%\liuye{list the invariant templates that \tool uses.}
The effectiveness of dynamic invariant detection largely depends on the diversity and scale of the customized invariant templates used.
In our methodology, these invariant templates are required to conform to the invariant specification language outlined in~\cref{fig:speclanguage}.
However, it is both impossible and impractical to cover every conceivable invariant template.
Our approach, akin to that of \invcon~\cite{Liu2022Invcon}, limits the scope to unary, binary, and ternary invariant templates.
Unlike \invcon, our templates are specifically designed for Solidity smart contracts, which are predominantly used for financial applications.
These contracts often entail intricate scientific computations on scalar variables.
%\liuye{and then how we customized our invariant templates}
Furthermore, Solidity features an array of complex data structures, such as \emph{mapping} and \emph{struct}.
To effectively infer invariants related to these structures, we have incorporated several derivation templates, such as \emph{MemberItem} and \emph{MappingItem}, which facilitate access to elements within these data structures.
Additionally, drawing inspiration from the significance of balance invariants as highlighted by
Wang et al.~\cite{Wang2019VUL}, we have introduced a \emph{SumMap} derivation template. This
template is specifically designed to aggregate the values contained within a mapping variable.

%\liuye{List all the templates we use in \tool?}

%<<<<<<< HEAD
%Invariant detection employs a statistical methodology to generate probable invariants with a certain degree of statistical confidence.
%Contrasting with the approach of \name{InvCon}, our method retains invariants that are refuted by certain transactions in the final results.
%This is because less stringent forms of these invariants, expressed as implications, may still hold true for certain contracts.
%=======
Dynamic invariant detection employs a statistical methodology to generate \emph{likely} primitive invariants with a certain degree of statistical confidence.
Contrasting with the approach of \invcon, our method retains invariants that are refuted by certain transactions in the final results.
This is because less stringent forms of these invariants, expressed as implications, may still hold true for certain contracts.
%>>>>>>> a544031 (minor)
Both the likely invariants and the falsified ones constitute a high-quality set of primitive predicates. Each of these predicates has been empirically verified through historical transaction data of smart contracts.
In our evaluation setting, each valid primitive invariant must be supported by at least three historical transactions.

\subsection{Modular Invariant Verifier}
The \houdini algorithm~\cite{flanagan2001houdini} is a widely recognized technique commonly used in
program annotation and validation processes.
Its primary objective is to automatically generate invariant annotations from a group of candidates.
%In our study, we utilize the \name{Houdini} algorithm to verify the authenticity of the invariants that we have extracted. Specifically, we focus on transforming implication invariant predicates into their equivalent logical forms, a process guided by De Morgan's Law:
%\[
%	A \implies B \equiv \lnot A \lor  B
%\]
To adapt \houdini algorithm for Solidity contracts, we initially instrument the contracts with the mined invariants.
This entails converting the invariants into a compatible format and then embedding them into the contract. The annotations are strategically placed at the beginning of functions to align with their specific names and arguments.
Subsequently, we transform the instrumented contracts into \emph{Boogie}~\cite{barnett2006boogie}
programs, leveraging the existing formal verification tool \verisol~\cite{wang2018formal} for
Solidity smart contracts.
We have refined the Boogie translator in \verisol to better accommodate contracts with
inheritance and polymorphism features.
For instance, the original translator lacked support for unnamed parent contract calls using the ``super'' keyword in Solidity, and it did not handle function overloading where a contract includes multiple functions with the same name.
We have enhanced its translation rules to effectively translate these complex contracts into {Boogie} programs.
Finally, we utilize {Boogie}'s own \houdini modular verifier to infer among the aforementioned invariant annotations, resulting in a set of verified invariants.

%\subsection{Implication Learner}
%\liuye{to check}
%The goal of the implication learner is to generate a useful set of implication candidates. Our preference leans towards implication invariants that have been substantiated by historical transactions.
%While this approach might lead to an incomplete set of implication invariants, it significantly reduces the number of impractical candidates, thereby accelerating the process as outlined in~\cref{algo:houdini}.
%As a result, the invariants refuted by our detection method must exhibit a minimum level of support in historical transaction data.
%In our evaluation framework, each valid implication invariant that is partially supported must be corroborated by at least three historical transactions.
%Then, the likely implication invariants will be verified by \name{Houdini}.
%The process will continue until no more implication candidates are generated.

\subsection{Suppressor}
An invariant is deemed redundant if it can be derived from another invariant.
The invariants verified by \tool may contain such redundancies.
Instead of eliminating redundancies in the dynamically detected invariants (as what
Daikon~\cite{daikon} does), we only remove redundancies from the invariants that are successfully
verified.
This design leaves more choices to the implication learner, when synthesizing implication
invariants.
%In contrast to Daikon~\cite{daikon}, which \textcolor{purple}{incorporates Simplify theorem
%prover}~\cite{detlefs2005simplify} to eliminate redundant likely invariant candidates derived
%from limited yet high-quality test cases crafted by developers, our tool, \tool, does not follow
%this practice for smart contracts.
%\yi{don't quite understand how Daikon eliminate?}
%%\liuye{added Simplify theorem prover description}
%Consequently, in \tool, redundant likely invariant candidates may emerge, leading to redundant
%verified invariants in the results.
%This design choice is motivated by the uncertainty regarding the extent to how sufficiently
%contracts have been ``tested'' in real-world transaction histories, which involve users of varying
%expertise.
Among the verified invariants, we utilize the Z3 solver~\cite{z3} to determine if one invariant
predicate can be deduced from another.
Following this analysis, we retain only the strongest invariant predicates in our final results.

%\liuye{dependency analysis technique?}

%The strategy to perform implication learning.
%
%what dependency analysis technique is used, slither..
%
%what constraint solving z3
%
%Constraint solving using Z3.

\section{Evaluation}\label{sec:eval}

In this section, we evaluate \tool to answer the following research questions:
\begin{enumerate}
	\item \textbf{RQ1:} How effectively does \tool generate invariants for smart contracts?
%	\item \textbf{RQ2:} How efficient is \tool in invariant generation for smart contracts?
	\item \textbf{RQ2:} How does the length of transaction histories used affect the performance of
	\tool?
	\item \textbf{RQ3:} How effective are the invariants detected by \tool in preventing real-world
	security attacks?
\end{enumerate}

\subsection{Methodology}

\paragraph{Benchmark}
To answer RQ1 and RQ2, we collected real-world smart contracts implementing the most popular \erc
and ERC721 standards, which have been studied extensively in previous
works~\cite{chen2019tokenscope,moon2022conformance,li2020securing,hildenbrandt2018kevm,li2019formal}.
The most important reason of choosing smart contracts implementing common standards is that their
invariant specifications are better understood, making it easier to obtain the ground truth.
First, we queried the public Ethereum ETL dataset hosted on the Google BigQuery
platform~\cite{etlbigquery} and then identified 13,116 contract addresses flagged as ERC20
deployed between 2021 and 2022.
Then, we identified 2,689 ERC721 contract addresses deployed between 2020 and 2022.
To facilitate our analysis, we kept only open-source contracts written in Solidity versions ranging
between 0.5.0 and 0.5.17, which are currently supported by \name{VeriSol}.
Finally, we obtained \ERCNumber ERC20 contracts and \NFTNumber ERC721 contracts for the
experimental evaluation, where each contract has at least 50 historical transactions as of June
2023.

%\paragraph{ERC20 and ERC721 Ground truth}
% \liuye{TODO: update}
To establish the ground truth for ERC20 and ERC721 contract specifications and ensure the included
invariants are comprehensive, we investigated multiple external sources.
These include the formal specifications referenced in the existing
literature~\cite{moon2022conformance,li2020securing}, popular smart contract libraries, such as
OpenZeppelin~\cite{OpenZeppelin}, and online documentations provided by smart contract formal
verification companies.
We list the collected ERC20 and ERC721 invariant specifications in \cref{tab:erc20-token-invs} and
\cref{tab:erc721-token-invs}, respectively.
These specifications are mainly based on
Certora~\cite{certora-erc20,certora-erc721,certora-pausable},
KEVM~\cite{rosu-erc20-k-2023,hildenbrandt2018kevm}, and OpenZeppelin API
documentations~\cite{openzeppelin-erc20-api,openzeppelin-erc721-api}.
We analyzed each of the collected invariants and manually translated it into our own specification
language (C.f.~\cref{fig:speclanguage}), which is a straightforward exercise in most cases.
We categorized these invariant specification into contract invariants, function preconditions and
postconditions in \cref{tab:erc20-token-invs,tab:erc721-token-invs}.
The functions listed in each table are the most commonly used standard functions for \erc
and ERC721 contracts.
%Because our invariants express only for contract invariant and function pre/postconditions,
Some specifications documented in external sources were omitted, e.g., ``Emitting a \code{Transfer}
event'' for the \code{transfer} function, because the particular language features are not
supported by our specification language.

\begin{table*}[t]
	\caption{Common ERC20 Invariants.}
	\label{tab:erc20-token-invs}
	\centering
	\scriptsize
	\begin{tabular}{p{2.5cm}|p{6cm}|p{8cm}}
		\toprule
		\textbf{Categories} & \textbf{Preconditions} & \textbf{Postconditions} \\
		\midrule
		transfer(to, amount)
		&
		\RaggedRight{
			\textbf{[a1]} msg.sender $\neq$ 0 \newline
			\textbf{[a2]} to $\neq$ address(0) \newline
			\textbf{[a3]} amount $\geq$ 0 \newline
			\textbf{[a4]} amount $\leq$ balances[msg.sender] \newline
			\textbf{[a5]} balances[to] + amount $\leq$ MAXVALUE
		}
		&
	 	\RaggedRight{
	 		\textbf{[b1]} to $\neq$ msg.sender $\implies$ balances[msg.sender] = old(balances[msg.sender]) - amount \newline
	 		\textbf{[b2]} to $\neq$ msg.sender $\implies$ balances[to] = old(balances[to]) + amount  \newline
	 		\textbf{[b3]} to = msg.sender $\implies$ balances[to] = old(balances[to]) \newline
	 		\textbf{[b4]} to = msg.sender $\implies$ balances[msg.sender] = old(balances[msg.sender]) \newline
			\textbf{[b5]} totalSupply = old(totalSupply)
		}\\
		\midrule
		transferFrom (from, to, amount)
		&
		\RaggedRight{
			\textbf{[a6]} from $\neq$ address(0) \newline
			\textbf{[a7]} to $\neq$ address(0) \newline
			\textbf{[a8]} amount $\geq$ 0 \newline
			\textbf{[a9]} amount $\leq$ balances[from] \newline
			\textbf{[a10]} amt $\leq$ allowed[from][msg.sender]  \newline
			\textbf{[a11]} balances[to] + amount $\leq$ MAXVALUE
		}
		&
		\RaggedRight{
			\textbf{[b6]} allowed[from][msg.sender] = old(allowed[from][msg.sender]) - amount  \newline
			\textbf{[b7]} from $\neq$ to $\implies$ balances[from] = old(balances[from]) - amount   \newline
			\textbf{[b8]} from $\neq$ to $\implies$ balances[to] = old(balances[to]) + amount \newline
		    \textbf{[b9]} from = to $\implies$ balances[from] = old(balances[from]) \newline
			\textbf{[b10]} allowed[from][msg.sender] = old(allowed[from][msg.sender]) - amount \newline
			\textbf{[b11]} totalSupply = old(totalSupply)
		}\\
		\midrule
		 approve(spender, amount)
		&
		\RaggedRight{
			\textbf{[a12]} amount $\geq$ 0 \newline
			\textbf{[a13]} spender $\neq$ address(0)
		}
		&
		\RaggedRight{
			\textbf{[b12]} allowed[msg.sender][spender] = amount  \newline
			\textbf{[b13]} totalSupply = old(totalSupply)
		}\\
		\midrule
		\RaggedRight{
			 increaseAllowance(\newline spender, amount)
		 }
		&
		\RaggedRight{
			\textbf{[a14]} spender $\neq$ address(0) \newline
			\textbf{[a15]} amount $\geq$ 0 \newline
			\textbf{[a16]} allowed[msg.sender][spender] + amount $\leq$ MAXVALUE
		}
		&
		\RaggedRight{
			\textbf{[b14]} allowed[msg.sender][spender] = old(allowed[msg.sender][spender]) + amount   \newline
			\textbf{[b15]} totalSupply = old(totalSupply)
		}\\
			\midrule
			\RaggedRight{
			decreaseAllowance(\newline spender, amount)
		}
		&
		\RaggedRight{
			\textbf{[a17]} spender $\neq$ address(0) \newline
			\textbf{[a18]} amount $\geq$ 0 \newline
			\textbf{[a19]} allowed[msg.sender][spender] $\geq$ amount
		}
		&
		\RaggedRight{
			\textbf{[b16]} allowed[msg.sender][spender] = old(allowed[msg.sender][spender]) - amount   \newline
			\textbf{[b17]} totalSupply = old(totalSupply)
		}\\
			\midrule
		mint(account, amount)
		&
		\RaggedRight{
			\textbf{[a20]} account $\neq$ address(0) \newline
			\textbf{[a21]} amount $\geq$ 0 \newline
			\textbf{[a22]} balances[account] + amount $\leq$ MAXVALUE
		}
		&
		\RaggedRight{
			\textbf{[b18]} balances[account] = old(balances[account]) + amount  \newline
			\textbf{[b19]} totalSupply = old(totalSupply) + amount
		}\\
		\midrule
		burn(from, amount)
		&
		\RaggedRight{
			\textbf{[a23]} from $\neq$ address(0) \newline
			\textbf{[a24]} amount $\geq$ 0 \newline
			\textbf{[a25]} balances[from] $\geq$ amount
		}
		&
		\RaggedRight{
			\textbf{[b20]} balances[from] = old(balances[from]) - amount  \newline
			\textbf{[b21]} totalSupply = old(totalSupply) + amount
		}\\
			\midrule
		pause()
		&
		\RaggedRight{
			\textbf{[a26]} paused = false
		}
		&
		\RaggedRight{
			\textbf{[b22]} paused = true
		}\\
		\midrule
		unpause()
		&
		\RaggedRight{
			\textbf{[a27]} paused = true
		}
		&
		\RaggedRight{
			\textbf{[b23]} paused = false
		}\\
			\midrule
		Contract Invariant
		&
		\multicolumn{2}{c}{
			\textbf{[c1]} totalSupply = SumMap(balances)
	}\\
		\bottomrule
	\end{tabular}
\end{table*}

\begin{table*}[t]
	\caption{Common ERC721 Invariants.}
	\label{tab:erc721-token-invs}
	\centering
	\scriptsize
	\begin{tabular}{p{2.5cm}|p{6cm}|p{8cm}}
		\toprule
		\textbf{Categories} & \textbf{Preconditions} & \textbf{Postconditions} \\
		\midrule
		(safe)-transferFrom(from, to, tokenId)
		&
		\RaggedRight{
			\textbf{[a28]} from = \_tokenOwner[tokenId] \newline
			\textbf{[a29]} from $\neq$ address(0) \newline
			\textbf{[a30]} to $\neq$ address(0) \newline
			\textbf{[a31]} (msg.sender = from $\lor$  msg.sender = \newline \_tokenApprovals[tokenId] $\lor$ \newline \_operatorApprovals[from][msg.sender]  = true)
		}
		&
		\RaggedRight{
			\textbf{[b24]} from $\neq$ to $\implies$ \_ownedTokensCount[from] = old( \_ownedTokensCount[from]) - 1  \newline
			\textbf{[b25]} from $\neq$ to $\implies$ \_ownedTokensCount[to] = old( \_ownedTokensCount[to]) + 1 \newline
			\textbf{[b26]} from = to $\implies$ \_ownedTokensCount[from] = old( \_ownedTokensCount[from]) \newline
			\textbf{[b27]} from = to $\implies$ \_ownedTokensCount[to] = old( \_ownedTokensCount[to]) \newline
			\textbf{[b28]} \_tokenOwner[tokenId] = to \newline
			\textbf{[b29]} \_tokenApprovals[tokenId] = address(0)
		}\\
		\midrule
		approve(to, tokenId)
		&
		\RaggedRight{
			\textbf{[a32]} \_tokenOwner[tokenId]  $\neq$ address(0)  \newline
			\textbf{[a33]} (msg.sender = \_tokenOwner[tokenId] $\lor$ \newline \_operatorApprovals[\_tokenOwner[tokenId] ][msg.sender]  = true)
		}
		&
		\RaggedRight{
			\textbf{[b30]} \_tokenApprovals[tokenId] = to
		}\\
		\midrule
		\RaggedRight{
			setApproveForAll(\newline operator, \_approved)}
		&
		\RaggedRight{
			\textbf{[a34]} operator $\neq$ msg.sender
		}
		&
		\RaggedRight{
			\textbf{[b31]} \_operatorApprovals[msg.sender][operator]  = \_approved
		}\\
		\midrule
		Contract Invariant
		&
		\multicolumn{2}{c}{
			\textbf{[c2]} len(\_tokenOwner) = SumMap(\_ownerTokenCount)
		}\\

		\bottomrule
	\end{tabular}
\end{table*}

\paragraph{Evaluation Metrics}
%Because there lacks a universal way to evaluate the quality of the generated invariants,
%To evaluate the effectiveness of \tool,
We use two evaluation metrics to evaluate \tool on the ERC20 and ERC721 contracts. % for \erc
%consitency checking compared with the ground truth listed in~\cref{tab:erc20-token-invs}.
Particularly, we use \textbf{Precision} and {${\textbf{Recall}}_{\textbf{ERC20}}$}
({${\textbf{Recall}}_{\textbf{ERC721}}$}) to measure the effectiveness of the generated invariants,
denoted as $X_{proved}$.
We denote the ground truth invariants (e.g., ERC20) as $Y$.
Formally,
\begin{align}
	&\textbf{Precision} = \frac{|X_{proved}|} {|X|},   \label{eq:precision}  \\
    {\textbf{Recall}}_{\textbf{ERC20}} & ({\textbf{Recall}}_{\textbf{ERC721}}) = \frac{|X_{proved}
    \cap Y|}{|Y|},  \label{eq:recall}
\end{align}
where \emph{precision} refers to the proportion of the generated invariants which are correct and
\emph{recall} is the proportion of the ground truth invariants which can be successfully generated.
Since the contract execution  trace set $\Delta$ from transaction histories may only contain a subset of functions invocations,
i.e., some functions are never invoked.
For a fair comparison, let $Y \downharpoonright \Delta$ represent the ground
truth invariants for the functions appeared in the histories, and we use the adjusted recall in our
experiments: $\frac{|X_{proved} \cap Y|}{|Y \downharpoonright \Delta|}$.

Note that although the ground truth invariants are derived based on multiple external sources and
widely deemed to be standard, they may still be incomplete, as there are infinitely many correct
invariants in theory.
The purpose of collecting the ground truth invariants is to include the list of common expectations
that are needed for contract safety and reliability.
On the other hand, certain smart contracts may not faithfully implement the ERC standards, and as a
result, either some ground-truth invariants may not hold for them or they satisfy additional
invariants not included in the ground truth.
Nevertheless, an ideal invariant generation tool should be able to recover as many ground-truth
invariants as possible, and meanwhile, recover other relevant invariants that are correct and
useful in describing unique smart contract behaviors.

%For further study, we also manually reviewed the correctness of the generated invariants that are not included in the ground truth.
%The review process is based on the following two criteria:
%To facilitate the review process, we check all the ERC721 contracts and sample 10 contracts out of \ERCNumber ERC20 contracts to study the resulting invariants, respectively.
%Two of the authors manually review the correctness of the generated invariants and a third author is consulted when there is a disagreement.
%

% \liuye{may need to update}
% Since in the limited transaction history, the invocation of some contract functions may not be observed.
% Except for contract-level invariants,
% the usefulness and recall are calculated based on the generated invariants and ground truth invariants of the observed functions.

\subsection{Experiment Setup}
%We measured the time usage of mining their invariants and identifying inconsistencies between the
%contract implementations and the ERC standard.
All the experiments were conducted on a desktop computer with the Ubuntu 20.04 OS, an Intel Core
Xeon 3.50GHx processor, and 32GB of RAM.
To facilitate the evaluation, we have crawled and cached all transaction histories in advance for
the contracts used in our experiments.

%\begin{table}
%	\small
%	\centering
%	\caption{Benchmarks.}
%	\begin{tabular}{lll}
%		\toprule
%		Category & \#Contracts & \#Txs  \\
%		\midrule
%		ERC-20 & 367 -\> 324&  51,643 \\
%		ERC721 & 23 -\> 10 & \\
%		\midrule
%		Overall & &  \\
%		\bottomrule
%	\end{tabular}
%\end{table}

\begin{table}[t]
	\small
	\centering
	\caption{The comparison results on ERC20 contracts.}
	\label{tab:comparison}
	\resizebox{\columnwidth}{!}{
	\begin{tabular}{lrrrr}\toprule
		\textbf{Tool} &\textbf{\#Inv} & \textbf{Prec.} &\textbf{Rec}$_{\textbf{ERC20}}$  &\textbf{Avg.time (s)} \\\midrule
		\invcon &413.23 & 0.095 &0.19  &13.99 \\
		\naive &480.89 & 0.094 &0.63 &15.25 \\
		\primitive &22.49 & 1.000 &0.61 &20.57 \\
		\tool &46.12 & 1.000 &0.80 &250.25 \\
		\bottomrule
	\end{tabular}
}
\end{table}

\subsection{RQ1: Effectiveness of Invariant Generation}
% We evaluate \tool in four scenarioes to investigate its effectiveness.  w/o $\mathcal{H}$ indicates no Houdnini verification, namely the original result by invariant detection, w/o $\mathcal{I}$ indicates no inference of implication invariants, namely the initial verification result produced by Houdini. w/o $\mathcal{P}$ represents partial invariant candiates that hold in the past transaction history are not applied.
% In contrast, \tool includes Houdini verification, implication invariant inference, and also apply partial invariant candidates.
\paragraph{Baseline}
To evaluate the performance of \tool, we used \textsc{InvCon} as our baseline.
\textsc{InvCon} uses Daikon as the back-end invariant detection engine
%where we implemented an intermediary input transformer that translates the execution traces of historic transactions to the compatible data trace files accepted by Daikon.
%In addition, we also added new derivation templates  into Daikon, \eg, MappingItem, to support some unique Solidity features.
and more implementation details can be found in the previous work~\cite{Liu2022Invcon}.
To the best of our knowledge, Cider~\cite{liu2022learning} is the only automated invariant
generation tool for smart contracts besides \invcon.
We have contacted the authors of Cider to obtain a copy of the
tool,\footnote{\url{https://github.com/UCSB-PLSE/Cider}} but failed to set it up.
We will discuss and compare with this work in~\cref{sec:relate}.

Additionally, we compared \tool with its two variants: \naive, which performs only
dynamic invariant detection tailored to Solidity contracts, and \primitive, which
employs the \houdini algorithm to generate verified invariants based only on dynamically
detected invariant candidates.

\paragraph{Results}
\Cref{tab:comparison} presents the comparison results for \ERCNumber ERC20 contracts, with a constraint of utilizing a maximum of 200 transactions per contract. The first column displays the names of the tools, while the second column enumerates the averaged number of invariants generated by each respective tool per contract. The middle two columns showcase the overall $\textbf{Precision}$ and $\textbf{Recall}_\textbf{ERC20}$ scores, and the last column provides the averaged time usage for each tool.

\tool achieves the highest recall score, reaching 0.80, and generates approximately 46 invariants
per contract, all of which are successfully verified by \verisol.
Notably, \invcon performs the least favorably in terms of the invariants generated, even
when compared with \naive.
Specifically, \invcon produces the second-highest number of invariants, yet its recall
score is significantly lower than that of \naive, while maintaining a similar
precision score of less than 0.1.
The poor performance is primarily attributed to the fact that \invcon's underlying
invariant detection engine, Daikon, supports only Boolean, integer/float, and string types native
to Java.
Consequently, the \textit{address} type (20 bytes long) in the Solidity language cannot be
seamlessly converted into a Java integer (8 bytes long).
Its conversion to the Java string type discards semantic information, rendering the straightforward
production of common invariants (e.g., \textbf{a1}, \textbf{a2}  in \cref{tab:erc20-token-invs})
unattainable for
\textsc{InvCon}.
Additionally, Daikon employs floating-point operations in arithmetic invariant templates (e.g.,
linear equation templates), which is not allowed in the Solidity semantics, leading to incorrect
invariants for \textbf{b6}, \textbf{b10} in \cref{tab:erc20-token-invs}.

\primitive exhibits a slightly lower recall score than \naive,
because some ground truth invariants that are inferred as likely invariants by
\naive may not be verified by \primitive.
This may be due to contract implementations slightly deviating from the standard.
For example, many ERC20 tokens do not enforce the precondition \textbf{a2} of the transfer function
in~\cref{tab:erc20-token-invs}, because transferring token to zero address could be used to
implement the token burning functionality.
The verified invariants are a more accurate reflection of the actual contract implementations,
compared with the likely invariants.
%likely invariants generated by \tool \textit{Naive} may not be correct.
%\yi{how would this explain? ground-truth invariants are all correct.}
Leveraging an algorithm capable of producing implications that widely exist in ERC20 invariants
(C.f. \cref{tab:erc20-token-invs}), \tool outperforms all the baseline tools, yielding 100\%
precise invariant results.
%\yi{why does \tool have a much better recall?}

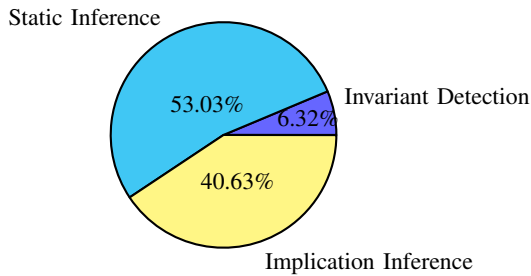
\begin{figure}[t]
	\centering
	\small
%	\scalebox{0.5}{
		\begin{tikzpicture}
			\pie[radius=1.5]{
				6.32/Invariant Detection,
				53.03/Static Inference,
				40.63/Implication Inference
			}
		\end{tikzpicture}
%	}
	\caption{The time usage by different components of \tool.}
	\label{fig:timeusage}
\end{figure}

Regarding the time usage, it is unsurprising that \tool takes the most time, whereas
\invcon and \naive finish the fastest.
In our experiments, we observed that the static inference process consumes the majority of the
time, constituting nearly 53\% of the overall time usage, as depicted in~\cref{fig:timeusage}.
This is primarily due to the iterative application of static inference
until no more implication candidates are provided.
Moreover, implication invariants generated in later iterations tend to be more intricate, resulting
in more complicated SMT formulas which take more time to solve.
To enhance the efficiency of \tool, we recommend capping the iterations used in the verification
process to four;
%considering that smart contracts are typically designed to be gas-efficient and
%bug-free \yi{what?}.
%For instance, most ERC20 or ERC721 contracts lack nested branch conditions. \yi{why does this
%matter for the cap?}
under such a setting, \tool demonstrates an averaged time savings of one minute in the entire
verification process without compromising the quality of resulting invariants.

%We also study the resulting invariants that are not included in the ground truth invariants.
%We sample 10 ERC20 contracts and manually review the {correctness} of the generated invariants.
%Both \tool and \textit{Houdini-Only} achieve a 100\% {correctness} score which means all the generated invariants are correct.
%In contrast, the invariant detection results by Daikon and \textit{Detection-Only} achieve only~\DAIKONCORRECT and~\DETECTIONCORRECT.

%\paragraph{Case Study}

Additionally, we investigated further on the contracts for which \tool generated additional
invariants deviating from the ground-truth ones.
Many of these contracts are found to be non-compliant with ERC20 specifications.
As illustrated in \cref{fig:tokenminterc}, we examined a real-world contract,
\code{TokenMintERC20Token}\footnote{\url{https://etherscan.io/address/0x62c23c5f75940c2275dd3cb9300289dd30992e59}},
where the \code{\_mint} function deviates from the contract invariant \textbf{c1}---the sum of
account balances always equals to the total supply---indicating non-compliance with the standard.
This discrepancy arises because only 1\% of the total supply tokens have been distributed to the
\code{Account} (Line~\ref{line:divide}).
\begin{figure}[t]
	\begin{minted}[fontsize=\scriptsize, escapeinside=||, highlightlines={9}]{solidity}
/** @dev Creates `amount` tokens and assigns them to `account`,
*   increasing the total supply.
* Requirements
* - `to` cannot be the zero address.*/
function _mint(address account, uint256 amount) internal {
 require(account != address(0), "ERC20: mint to the zero address");
 _totalSupply = _totalSupply.add(amount);
 _balances[account] = _balances[account].add(amount);
 _balances[Account] = _totalSupply/100; |\label{line:divide}|
}
	\end{minted}
	\caption{\code{TokenMintERC20Token} contract violating \textbf{c1}.}
	\label{fig:tokenminterc}
\end{figure}
%
%Furthermore, we also investigated on the invariants expected from the \erc standard, which \tool
%failed to detected.
%We found 16 contracts inconsistent with the \erc specification: six of them violate Inv\#1, one
%violates Inv\#4, and the remaining ones violate the invariants on \textit{transfer} and
%\textit{transferFrom}.

\begin{table}[t]
	\small
	\centering
	\caption{The mutation testing results on  ERC20 contracts against the verified invariants by \tool.}
	\label{tab:mutation}
\resizebox{\columnwidth}{!}{
	\begin{tabular}{p{3.3cm}rrr}\toprule
		\textbf{Categories} & \textbf{approve} & \textbf{transfer} & \textbf{transferFrom} \\
		\midrule
		No. total mutants & 1,539 & 1,141 & 297 \\
		No. killed mutants & 998 (64.8\;\%)& 624 (54.6\;\%)& 101 (34.0\;\%)\\
		\midrule
		P1. Contract invariants & 245 (24.5\;\%) & 344 (55.1\;\%)& 55 (54.4\;\%)\\
		P2. Function pre/post & 763 (76.4\;\%) & 465 (74.5\;\%)& 61 (60.4\;\%)\\
		\midrule
		P3. ERC20 standard   & 751 (75.2\;\%)& 266 (42.6\;\%)& 43 (42.5\;\%)\\
		P4. Non-ERC20 standard   & 995 (99.7\;\%)& 601 (96.3\;\%)& 98 (97.0\;\%) \\
		\bottomrule
	\end{tabular}
}
\end{table}

\paragraph{Invariant Quality}
Furthermore, to assess the significance of the invariants generated by \tool, we conducted mutation
testing on the same benchmark and computed the corresponding mutation scores against these
invariants.
\Cref{tab:mutation} presents the mutation testing results on ERC20 contracts,
specifically focusing on the three most important functions: \textit{approve}, \textit{transfer},
and \textit{transferFrom}.
We introduced six mutation operators, such as \textit{binary/unary-op-mutation} and
\textit{require-mutation}, along with the others, based on the mutation generator
Gambit~\cite{gambit} developed by Certora.\footnote{https://www.certora.com/}
This mutation-based approach was also adopted by Certora to evaluate the quality of smart contract
specifications.\footnote{https://docs.certora.com/en/latest/docs/gambit/index.html}
In total, we generated $1,539$, $1,141$, and $297$ mutants for \textit{approve}, \textit{transfer},
and \textit{transferFrom}, respectively.
\cref{tab:mutation} shows that 64.8\;\%, 54.6\;\%, and 34.0\;\% mutants of approve, transfer, and transferFrom are successfully killed, respectively.

To delve into those killed mutants, in \cref{tab:mutation}, we use P1, P2, P3, and P4 to denote different types of invariants and the corresponding rows show the number of killed mutants by these invariants.
Although the contract invariants (P1) accounts for 24.5\;\% to 54,4\;\% of the killed mutants,  function pre/post-conditions (P2) demonstrate a more substantial impact occupying at most 76.4\;\% of the killed mutants.
Moreover, non-ERC20 standard invariants successfully eliminate nearly the entire set (96\;\% more) of the total killed mutants.
In contrast, ERC20 standard invariants eliminate a smaller set of mutants.
This suggests that the invariants generated by \tool capture richer program semantics, contributing
to a more comprehensive set of invariant specifications for smart contracts.

Interestingly, \cref{tab:mutation} reveals that only 34\% of mutants related to the
\textit{transferFrom} function are successfully eliminated.
Upon investigation, we discovered that \textit{transferFrom} is overprotected, where one of its
function-level preconditions is redundant.
\Cref{fig:overprotect} depicts a common implementation of \textit{transferFrom}, facilitating token
transfer on behalf of the token owner through two internal functions, \textit{\_transfer} and
\textit{\_approve}.
This design rationale primarily aims at direct code reuse for the other two public functions,
\textit{transfer} and \textit{approve}.
However, in the \textit{transferFrom} function, both requirements (Line~\ref{line:req1} and
Line~\ref{line:req2}) check if the \textit{sender} parameter is a zero address.
Consequently, mutations on either Line~\ref{line:req1} or Line~\ref{line:req2} do not diminish the
requirements that \textit{transferFrom} should adhere to, resulting in a low mutation score for
\textit{transferFrom}.
It is noteworthy that redundant requirements in smart contracts lead to higher gas consumption
during transaction execution and should be minimized whenever possible.

\begin{figure}[t]
	\begin{minted}[escapeinside=||, highlightlines={2, 10}, fontsize=\scriptsize]{solidity}
function _transfer(address sender, address recipient, uint256 amount) internal {
 require(sender!=address(0), "zero address"); |\label{line:req1}|
 require(recipient!=address(0), "zero address");

 _balances[sender]=_balances[sender].sub(amount);
 _balances[recipient]=_balances[recipient].add(amount);
 emit Transfer(sender, recipient, amount);
}
function _approve(address owner, address spender, uint256 value) internal {
 require(owner!=address(0), "zero address"); |\label{line:req2}|
 require(spender!=address(0), "zero address");

 _allowances[owner][spender] = value;
 emit Approval(owner, spender, value);
}

function transferFrom(address sender, address recipient, uint256 amount) public returns (bool) {
 _transfer(sender, recipient, amount);
 _approve(sender, msg.sender, _allowances[sender][msg.sender].sub(amount));
 return true;
}
	\end{minted}
	\caption{Illustration of the overprotected \textit{transferFrom} function.}
	\label{fig:overprotect}
\end{figure}

%\begin{itemize}
%	\item  No. total mutants
%	\item  No. total killed mutants
%	\item  \textbf{P1} No. mutants killed by contract-level invariants
%	\item  \textbf{P2} No. mutants killed by function pre-/post invariants
%	\item  \textbf{P3} No. mutants killed by erc20 standard invariants
%	\item  \textbf{P4} No. mutants killed by non-erc20 standard invariants
%\end{itemize}

\begin{table}[t]
	\caption{ERC721 invariants generated by \tool.
		%		\textit{Note that the invariants (i.e., has prefix \textbf{a}, \textbf{b} and \textbf{c}) match with the ground truth in~\Cref{tab:erc721-token-invs}. \cx{Text too small}}
	}
	\label{tab:erc721-token-resulting-invs}
	\centering
	%	\scriptsize
	\resizebox{\columnwidth}{!}{
		\begin{tabular}{p{2cm}|p{2cm}|p{3cm}}
			\toprule
			\textbf{Category} & \textbf{Preconditions} & \textbf{Postconditions} \\
			\midrule
			(safe)-transferFrom
			&
			\RaggedRight{
				[a28, a29, a30]
			}
			&
			\RaggedRight{
				[b24, b25, b26, b27, b28, b29] % \newline
				%			[post1] \_ownedTokensCount[to] $\ge$ old(\_ownedTokensCount[to]) \newline
				%			[post2] \_ownedTokensCount[from] $\ge$ old(\_ownedTokensCount[from])
			}\\
			\midrule
			approve
			&
			\RaggedRight{
				[a32] % \newline
				%			msg.sender $\neq$ address(0) \newline
				%	 	\textcolor{gray}{	\_tokenOwner[tokenId]  $\neq$ address(0)}  \newline
				%			[pre1] \_tokenOwner[tokenId] $\neq$ to
			}
			&
			\RaggedRight{
				[b30]
			}\\
			\midrule
			setApproveForAll
			&
			\RaggedRight{
				%			msg.sender $\neq$ address(0) \newline
				[a34]
			}
			&
			\RaggedRight{
				[b31]
			}\\
			\midrule
			Contract Invariant
			&
			\multicolumn{2}{c}{
				[c2]
			}\\

			\bottomrule
		\end{tabular}
	}
\end{table}

\paragraph{Invariant Crowdsourcing}
Less popular smart contracts may have scarce transaction histories.
For example, many ERC721 contract instances may not have enough transactions to infer high-quality
invariants.
Each contract instance can slightly deviate from the standard specifications, therefore, we
hypothesize that reverse engineering invariants from a single contract and its limited transaction
histories is inferior to that from multiple contracts.
We validate this hypothesize on a set of \NFTNumber ERC721 contracts, restricting the evaluation
to at most 200 transactions per contract.
The objective is to examine \tool's effectiveness in recovering the ground truth invariants listed
in~\cref{tab:erc721-token-invs} by combining invariant results from multiple contracts.
Notably, to achieve meaningful combination,
every invariant result will be normalized according to a universal ERC721 definition on the name of state variables and the name of function input variables.

%The rationale behind this approach is that each contract
%has a diverse transaction history covering only a limited set of normal program behaviors, and
%combining these histories may contribute to a more comprehensive set of invariant results.
%\yi{This needs a bit more explanation: why does ERC721 need this approach? what's the implication
%in practice?}

\Cref{tab:erc721-token-resulting-invs} presents the combined invariant results from all ERC721
contracts. It demonstrates that \tool successfully recovers the contract invariants, all
postconditions, and nearly all preconditions except \textbf{a31} and \textbf{a33}, which contain
disjunctions over predicates. Consequently, the combination of invariant results from multiple
contracts significantly improves the overall recall rate (14/16).

\begin{tcolorbox}[size=title, opacityfill=0.1]
	\textbf{Answer to RQ1}:
	\tool is able to reverse engineer standard invariant specifications from contract transaction
	histories and takes no more than five minutes per contract. Additionally, the uncommon invariants
	generated for ERC20 contracts capture important program semantics beyond the established
	standards. Moreover, the evaluation on ERC721 contracts demonstrates the advantage to mine common
	invariants from multiple contracts and their transaction histories.
%\yi{how does it do for ERC20 and ERC721. need a bit more details.}
	%We conclude that \tool is accurate.
\end{tcolorbox}

\begin{figure}[t]
	\includegraphics[width=\columnwidth]{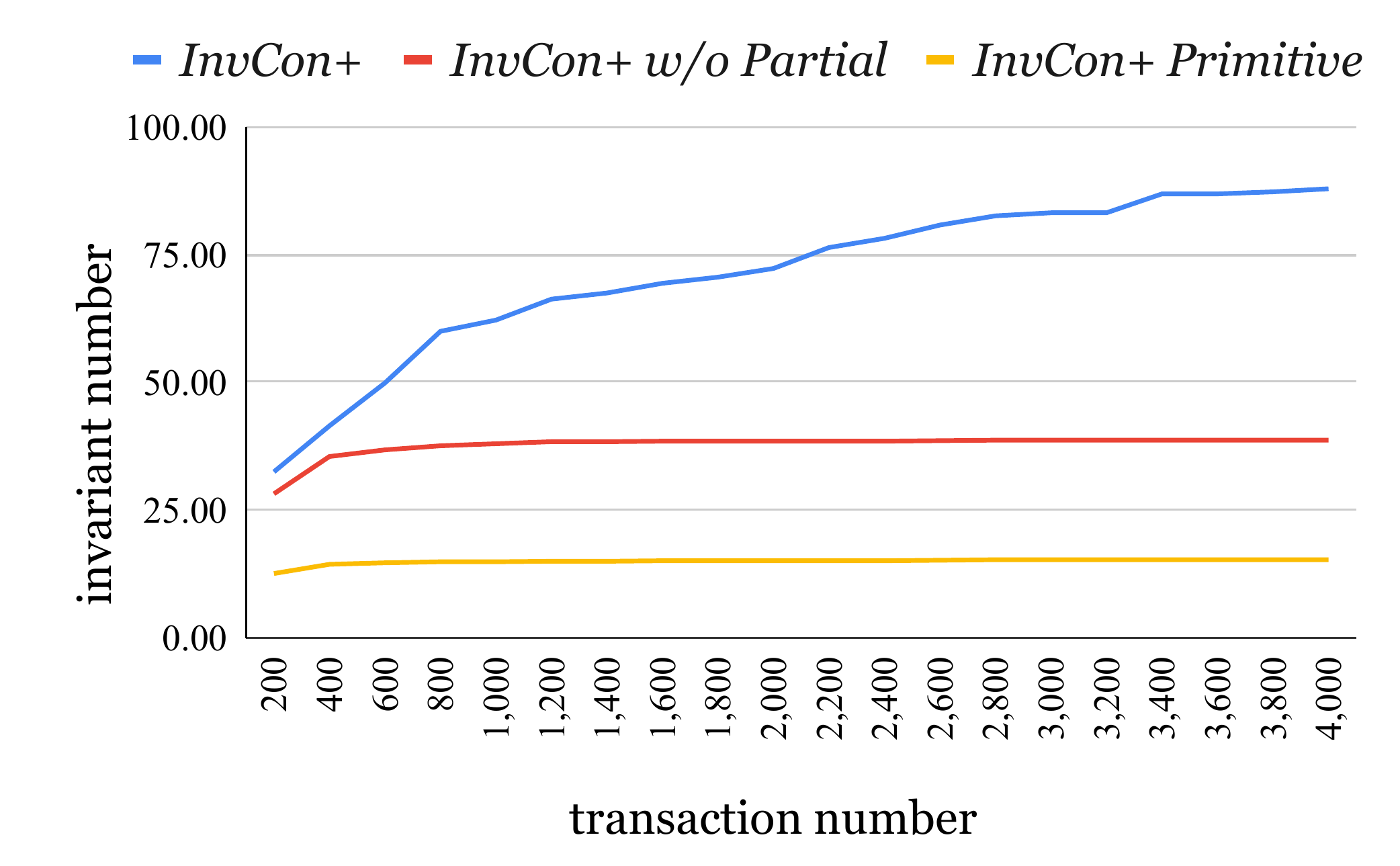}
	\caption{The averaged number of invariants generated with different number of transactions.}
	\label{fig:invnumber}
\end{figure}
% \begin{figure}
% 	\includegraphics[width=\columnwidth]{./figures/RQ3-Precision.pdf}
% 	\caption{The \textbf{precision*} score of the invariant results.}
% \end{figure}
\begin{figure}[t]
	\includegraphics[width=\columnwidth]{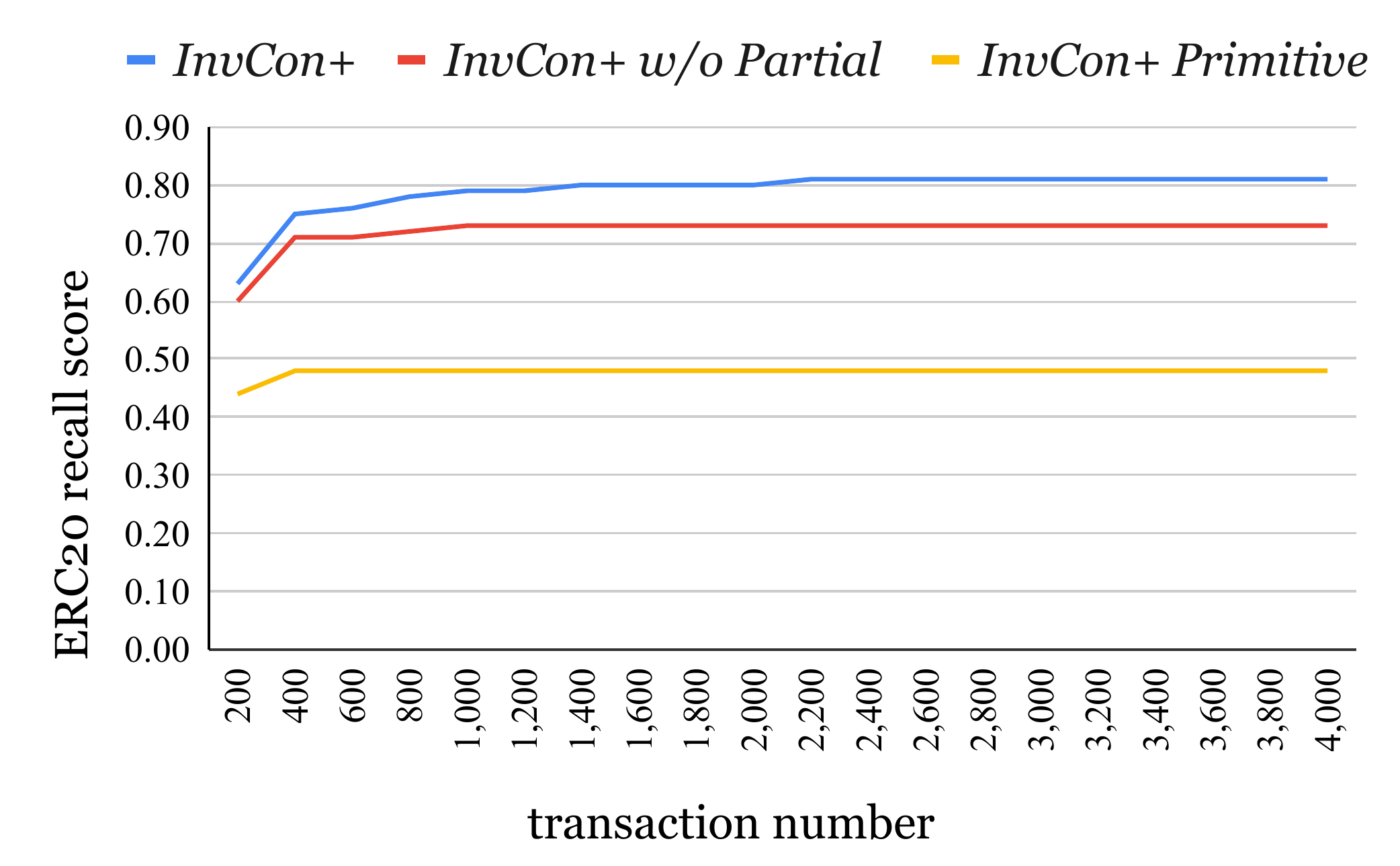}
	\caption{The averaged ERC20 recall score of the invariant results generated with different number of transactions. }
	\label{fig:recall}
\end{figure}
%\begin{table*}[t]
%	\centering
%	\scriptsize
%	\caption{Statistical result for the impact of varied transaction histories.}\label{tab:statistical}
%	\begin{tabular}{lrrr|rrr|rrr|ccc}\toprule
%		\multirow{2}{*}{Metric} &\multicolumn{3}{c|}{\tool} &\multicolumn{3}{c|}{\tool (w/o p.) \textcircled{2}} &\multicolumn{3}{c|}{Houdini-Only \textcircled{1}} &\multicolumn{3}{c}{$\hat{A_{12}}$} \\\cmidrule{2-13}
%		&Avg. &Var. &p-value &Avg. &Var. &p-value &Avg. &Var. &p-value &[\textcircled{2}, \textcircled{1}] &[\tool, \textcircled{1}] &[\tool, \textcircled{2}] \\\midrule
%	Avg.\#VerifiedInv &63.66 &748.85 &0.00 &28.53 &1.45 &0.00 &12.09 &0.61 &0.00 &1 &1.00 &0.99 \\
%	Precison &0.32 &0.03 &0.00 &0.58 &0.00 &0.00 &0.66 &0.00 &0.00 &0 &0.00 &0.01 \\
%	Recall &0.73 &0.00 &0.00 &0.51 &0.04 &0.00 &0.44 &0.01 &0.00 &0.945 &1.00 &0.99 \\
%		\bottomrule
%	\end{tabular}
%\end{table*}
%

\subsection{RQ2: Impact of Transaction Histories}

The length of the transaction histories used can influence the effectiveness of \tool.
To investigate this impact, we selected the top 10 ERC20 contracts with the longest transaction
histories, ensuring that all the chosen contracts have a history of at least 10,000 transactions.
In evaluating the influence of transaction history length, we employed the earliest 4,000
transactions and divided them into 20 groups, each subsequent group having 200 more transactions
than the previous one.

We utilized \primitive as the baseline and compared with it on the number of verified
invariants and the corresponding recall score.
Additionally, to explore the effect of applying the detected partially supported invariant
candidates, which hold for a subset of the transaction histories, we compared \tool with a
variant, \wopartial, that does not use these partial candidates.
In this experiment, we considered the ground truth invariants from the functions which are observed
in the earliest 4,000 transactions, when computing the recall score, i.e.,
${\textbf{Recall}}_{\textbf{ERC20}}$.

\Cref{fig:invnumber} illustrates the number of verified invariants per contract corresponding to the use of
different transaction history lengths. The impact of transaction history size on the number of
verified invariants is evident, with \tool generating the most invariants, followed by \tool
\wopartial.
This demonstrates that the partially supported invariant candidates, although do not hold on their
own, may be useful in constructing richer implication invariants.
By incorporating partial invariant candidates, \tool captures subtle contract behaviors more
effectively, resulting in more comprehensive invariant specifications---approximately two times and
one time more than \wopartial and \primitive, respectively.

In \cref{fig:recall}, the recall score of invariant results is presented for varying transaction
history lengths. Clearly, all recall scores increase with longer transaction histories, as more
function invocations are observed. Notably, \tool achieves a higher recall score compared to the
baselines. The figure also indicates a more significant gain in recall score from 200 to 400
transactions, with negligible gains after 400, 1,000, and 2,200 transactions for \primitive, \wopartial, and \tool, respectively.
This observed difference suggests that \tool has a higher chance of capturing more comprehensive
invariant specifications with increased transaction histories.
Additionally, to effectively apply \tool, it is recommended to use around 2,000 transactions for
invariant detection.

%\Cref{fig:invnumber} shows the number of verified invariants per contracts with different length of transaction histories.
%It is clear that the number of verified invariants are affected by the transaction history size.
%\tool generates the most verified invariants, followed by \tool \textit{w/o partial}.
%This indicates that by using partial invariant candidate, \tool can subsume as more subtle user behaviors as possible which can facilitate the production of more comprehensive invariant specifications, which are usually two times and one times more than \tool \textit{w/o partial} and \tool \textit{Houdini}, respectively.
%
%\Cref{fig:recall} shows the recall score of invariant results with different length of transaction histories.
%Clearly, all the recall score will increase with the longer transaction history.
%This is because more function invocations will be observed within a larger history.
%\tool achieves the higher recall score compared with others.
%\Cref{fig:recall} also shows that the recall score has a larger gain from 200 transactions to 400 transactions but negligible gain after 400, 1000, 2200 transactions for \tool \textit{Houdini}, \tool \textit{w/o partial}, and \tool, respectively.
%Such fluctuation difference implies that \tool has higher chance to capture comprehensive invariant specifications with more transaction histories.
%Also, to effectively apply \tool, we suggest it is sufficient enough to use around 2,000 transactions for smart contract invariant detection.

\begin{tcolorbox}[size=title, opacityfill=0.1]
	\textbf{Answer to RQ2}:
	The scale of transaction histories affect the invariant results of \tool, while longer histories
	empower \tool to generate more comprehensive invariant specifications.
	%We conclude that \tool is accurate.
\end{tcolorbox}

\subsection{RQ3: Application in Securing Smart Contracts}
The invariants generated by \tool capture the key semantics of smart contracts under normal executions, which may serve as a basis for formal contract specifications.
High-quality contract specifications have been shown to be effective in securing smart contracts through runtime validation~\cite{li2020securing} and static verification~\cite{wang2018formal}.
To answer RQ3, we evaluated \tool on a set of benchmark contracts from SECBIT~\cite{secbit2018buggyerc20}, which contains 25 types of vulnerabilities in real-world ERC20 contracts exposed to security attacks that have resulted in significant financial losses.
%\cx{Removed "labs" for name consistency with introduction.}

\Cref{tab:commonerc20vul} provides an overview of the verification results for the evaluated ERC20
contracts, categorized by vulnerability types. It contains information about the overall count of
vulnerabilities and the effectiveness of our generated invariants in detecting them.
The benchmark contracts used in our evaluation encompass 9 instances of integer overflow
vulnerabilities and 16 other vulnerability types.
However, certain vulnerabilities are beyond the scope of formal specifications, such as
\textbf{v14}, \textbf{v21}, and \textbf{v24} which are related to constructor naming, \textbf{v15}
and \textbf{v16} which are associated with different Solidity versions, and \textbf{v23} which
pertains to function visibility.
We focused on the remaining 18 types of vulnerabilities.
Note that some of the vulnerabilities identified are beyond the specifications outlined in the
ERC20 standard (see \cref{tab:erc20-token-invs}) and they can only be detected using richer
customized specifications.

For each of vulnerability types, we evaluated the verification results of the invariants generated
by \tool on the corresponding benchmark contracts.
We selected the top three contracts with the highest occurrence of each vulnerability type and
assessed whether the invariants detected by \tool could prevent the corresponding attacks on these
contracts.
The results are shown in \cref{tab:commonerc20vul}.
We found that \tool was able to detect all overflow vulnerabilities in the benchmark contracts.
For instance, \cref{fig:BeautyChain} demonstrated that \tool detected the integer overflow vulnerability (CVE-2018-10299) in the \textit{batchTransfer} function of the \code{BEC} contract.
This vulnerability is caused by the unchecked multiplication of \code{cnt} and \code{\_value} in
Line~\ref{line:bec1}.
If an attacker calls \code{batchTransfer} with a large \code{cnt} value, the
unsigned integer \code{amount} will overflow, potentially allowing the attacker to receive more
tokens than intended.
However, such a transaction would violate invariant \textbf{c1} in \cref{tab:erc20-token-invs}, as
the \code{totalSupply} would no longer equal to the sum of all balances.
Thus, such an attack can be effectively prevented, if the generated invariants are enforced for
each function execution.

%Moreover, \tool is capable of identifying the vulnerability in the \textit{transferProxy} functions of the \textit{SMT} contract, as illustrated in \Cref{fig:SmartMesh}. This vulnerability occurs due to an incorrect parameter passed to the \textit{ecrecover} function.
%The \textit{keccak256} function in Line~\ref{line:smt1} is used to generate the signature of the public key, and \textit{ecrecover} is then employed to retrieve the public key from the signature.
%When the parameter of \textit{ecrecover} is incorrect, it returns a zero address.
%An attacker can exploit this by using the zero address as the \textit{_from} parameter to call the \textit{transferProxy} function, circumventing the check in Line~\ref{line:smt2}, and thus transfer tokens without proper authorization.
%\tool can detect this vulnerability by proactively managing zero addresses to prevent their misuse as arguments.

\begin{figure}[t]
\small
\begin{minted}[escapeinside=||,texcomments, highlightlines={3}]{solidity}
function batchTransfer(address[] _receivers, uint256 _value) public whenNotPaused returns (bool) {
uint cnt = _receivers.length;
uint256 amount = uint256(cnt) * _value;|\label{line:bec1}|
require(cnt > 0 && cnt <= 20);
require(_value > 0 && balances[msg.sender] >= amount);

[msg.sender] = balances[msg.sender].sub(amount);
for (uint i = 0; i < cnt; i++) {
balances[_receivers[i]] = balances[_receivers[i]].add(_value);
Transfer(msg.sender, _receivers[i], _value);
}
return true;
}
\end{minted}
\caption{batchTransfer function in BEC contract.}
\label{fig:BeautyChain}
\end{figure}

\tool is unable to detect some remarkable mistakes that totally deviate from programmer expectations.
For example,
\textbf{v11} is a vulnerability that allows any party to halt the token transfer process. This
issue arises from the modification of the \code{onlyFromWallet} modifier, wherein ``\code{==}'' was
mistakenly replaced with ``\code{!=}''.
Consequently, anyone other than \code{walletAddress} can arbitrarily invoke the two permissioned
functions: \code{enableTokenTransfer} and \code{disableTokenTransfer}.
\tool failed to detect this vulnerability for two primary reasons.
First, the \code{onlyFromWallet} function is not specified in the ERC20 standard, preventing the
application of the existing invariant templates.
Second, the contract histories contain many irregular behaviors exploiting these functions,
hindering \tool from inferring correct invariants related to \code{onlyFromWallet}.

%Beyond the semantics of the ERC20 standard, \tool does not identify this vulnerability.

%\begin{figure}[t]
%\small
%\begin{minted}[escapeinside=||,texcomments]{solidity}
%function transferProxy(address _from, address _to, uint256 _value, uint256 _feeMesh,
%uint8 _v,bytes32 _r, bytes32 _s) public transferAllowed(_from) returns (bool){
%...
%bytes32 h = keccak256(_from,_to,_value,_feeMesh,nonce,name);|\label{line:smt1}|
%if(_from != ecrecover(h,_v,_r,_s)) revert();|\label{line:smt2}|
%...
%return true;
%}
%\end{minted}
%\caption{transferProxy function in SMT contract.}
%\label{fig:SmartMesh}
%\end{figure}
% Overall, the success rates for all cases are higher than {xxx} and the verification time is less than {xxx}.
% Specifically,
% {xxx} accepts the most given ERC20 invariants while {BEC} accepts the least.
% {BEC} is a buggy ERC20 contracts where its \textit{batchTransfer} function contains integer overflow vulnerability.
% {xxx} accepts the most given ERC721 invariants while {xxx} accepts the least.
% {xxx} spent the largest verification time while {xxx} spent the least.
% The reason is that {xxx} has {xxxx} and {xxxx}.
%In conclusion,
%\tool is able to detect real-world smart contract vulnerabilities and we believe \tool can be applied to ensure the security and reliability of smart contracts.

 \begin{tcolorbox}[size=title, opacityfill=0.1]
 	\textbf{Answer to RQ3}:
 \tool is able to detect invariants that are useful for preventing real-world smart contract
 vulnerabilities.
 Enforcing invariants in contract executions may ensure the security and reliability of smart
 contracts.
 	%We conclude that \tool is accurate.
 \end{tcolorbox}

 \begin{table}[t]
	\small
	\caption{Common ERC20 vulnerabilities.}
  \resizebox{\columnwidth}{!}{
	\begin{tabular}{l|p{4.5cm}|r|c}
		\toprule
		\textbf{ID} & \textbf{Vulnerability Types} &\textbf{Total} &\textbf{Detected} \\
		\midrule
		v1 & batchTransfer-overflow &13 &Yes \\
		v2 & totalsupply-overflow &521 &Yes \\
		v3 & verify-invalid-by-overflow &2 &Yes \\
		v4 & owner-control-sell-price-for-overflow &1 &Yes \\
		v5 & owner-overweight-token-by-overflow &9 &Yes \\
		v6 & owner-decrease-balance-by-mint-by-overflow &487 &Yes \\
		v7 & excess-allocation-by-overflow &1 &Yes \\
		v8 & excess-mint-token-by-overflow &9 &Yes \\
		v9 & excess-buy-token-by-overflow &4 &Yes \\
		\midrule
		v10 & verify-reverse-in-transferFrom &79 &Yes \\
		v11 & pauseTransfer-anyone &1 &No \\
		v12 & transferProxy-keccak256 &10 &Yes \\
		v13 & approveProxy-keccak256 &10 &Yes \\
		v14 & constructor-case-insensitive &4 &N/A \\
		v15 & custom-fallback-bypass-ds-auth &1 &N/A \\
		v16 & custom-call-abuse &144 &N/A \\
		v17 & setowner-anyone &3 &Yes \\
		v18 & allowAnyone &4 &Yes \\
		v19 & approve-with-balance-verify &18 &Yes \\
		% v20 & re-approve &0 &N/A \\
		v20 & check-effect-inconsistency &1 &Yes \\
		v21 & constructor-mistyping &4 &N/A \\
		v22 & fake-burn &2 &Yes \\
		v23 & getToken-anyone &3 &N/A \\
		v24 & constructor-naming-error &1 &N/A \\
		\bottomrule
	\end{tabular}}
	\label{tab:commonerc20vul}
\end{table}

\section{Related Work}
\label{sec:relate}

The related works can be broadly categorized into smart contract security analysis and invariant inference.
\subsection{Smart Contract Security Analysis}

%\paragraph{Outline}
%\begin{itemize}
%	\item Organize smart contract security analysis based on the vulnerabilities/root causes.
%	\item Organize smart contract security works based on their techniques.
%	\item List the challenging problems in securing smart contracts
%	\item Our work can mitigate the problems and complement existing security analysis works.
%\end{itemize}
The security analysis primarily focuses on detecting smart contract vulnerabilities.
Common vulnerabilities in smart contracts include integer overflow/underflow~\cite{BECoverflow}, reentrancy~\cite{DAOattacks}, and dangerous delegatecall operations~\cite{Paritystolen}.
For instance, in 2017, the Parity wallet contract was hacked due to missing protection for the delegatecall operation, a feature that allows one contract to securely delegate part of its functionality to another contract. As a result, the attacker gained control of the wallet and stole 150,000 ETH, valued at approximately \$30 million USD at the time.

These common vulnerabilities have been extensively studied in~\cite{wang2020oracle, feist2019slither, brent2020ethainter, tikhomirov2018smartcheck, securify, feng2019precise, luu2016making, manticore, mythril, kalra2018zeus, 2018contractfuzzer, wustholz2020harvey, echidna}.
Most static analysis tools, such as Slither~\cite{slither}, Securify~\cite{securify}, Zeus~\cite{kalra2018zeus}, and Ethainter~\cite{brent2020ethainter}, utilize control-flow, data-flow, or taint-flow analysis for vulnerability detection, usually achieving a high recall but low precision rate.
In contrast, the others~\cite{ manticore, mythril, oyente} use symbolic execution for program path exploration to identify contract vulnerabilities, along with a higher precision but lower recall rate.
There are also formal verification tools for ensuring the correctness of functional properties~\cite{so2020verismart, hajdu2020solc, permenev2020verx}, and workflow policy~\cite{wang2018formal} in smart contracts.
%Static verification~\cite{wang2018formal} and theorem proving~\cite{hirai2017defining} are also applied for smart contract security.
The dynamic analyses~\cite{2018contractfuzzer, echidna, Wang2020OSD, nguyen2020sfuzz, Liu2022FPB} perform random or model-based testing on smart contracts and then check execution result against predefined oracles for finding a wide range of vulnerabilities.
Although these tools have been proven effective in detecting common vulnerabilities,
%their capability is largely restricted by the usage of simple and general oracles.
unfortunately, Zhang et al.~\cite{zhang2023demystifying} found that only 20.5\% of real-world smart contract bugs can be successfully detected by state-of-the-art tools.
This is because the existing tools use simple, generic, and hard-coded security patterns or oracles, which are ineffective to recognize subtle logic bugs on specific contracts.

Because there is no one-for-all patterns or oracles for identifying contract logic bugs, most valued Web3 projects hire third-party security auditing companies to manually review their contracts.
Despite undergoing costly code auditing, numerous projects still fall victim to security breaches~\cite{hacks}.
In our opinion, one root cause is that contract developers and the corresponding auditors may have
divergent expectations on smart contracts, which are not easy to pinpoint without sufficient
contract specifications.
Therefore, apart from enhancing existing security tools, the invariants generated by \tool can
reinforce contract specifications to mitigate the incompleteness and inaccuracy issues of automated
verification and contract auditing.
%There is a large body of work on smart contract analysis~\cite{ durieux2020empirical}.
%%\Fix{As shown in an empirical study~\cite{ferreira2020smartbugs, durieux2020empirical}},
%Most focus on the security issues of smart contracts via static and dynamic analyses combined with a set of predefined vulnerability patterns.
%Slither~\cite{slither} is a static analyzer which runs a suite of more than 76 vulnerability detectors to find smart contract security bugs.
%Oyente~\cite{oyente} is one of the earliest symbolic execution engine, which detects eight contract vulnerabilities.
%The other symbolic execution-based tools also include Manticore~\cite{manticore} and Mythril~\cite{mythril}, where Manticore is able to detect 11 vulnerabilities and Mythril can find 36 vulnerabilities.
%As for the dynamic analysis tools,
%ContractFuzzer~\cite{2018contractfuzzer} is the earliest dynamic fuzz testing tool targeting common vulnerability types, such as reentrancy,
%followed by ContraMaster~\cite{Wang2020OSD}, sFuzz~\cite{nguyen2020sfuzz}, and SPCon for permission bug detection~\cite{Liu2022FPB}.
%\tool complements the existing tools by inferring likely invariants, which can then be used to
%augment their oracles by strengthening contract specifications.
%\liuye{For example, Inv\#1 :``$\sum_u$ balances[u] = \_totalsupply'' can be used to capture
%integer underflow/overflow of \erc contracts.}
%\yi{unclear how and what security patterns}

\subsection{Invariant Inference}
%Check this paper "Learning Contract Invariants Using Reinforcement Learning"
%\url{https://dl.acm.org/doi/pdf/10.1145/3551349.3556962}
%\paragraph{Outline}
%\begin{itemize}
%	\item The common goal and techniques of specification mining.
%	\item The well-known specification mining works: dynamic invariant detection (Daikon) and static specification mining/invariant detection (xxx).
%	\item List existing related works on the specification mining of smart contracts. (Deep-learning based invariant detection for smart contracts. Need to check.)
%	\item Discuss how our work distinguish itself from them.
%\end{itemize}

%Static and dynamic invariant detection are two fundamental approaches in program analysis, each with its unique methods and applications.
The static and dynamic invariant inference have been well-studied for traditional programs.
ESC/Java~\cite{flanagan2002extended} is a well-known static checking tool for Java programs.
It leverages invariant annotations to define properties in the code, improving the precision of static checking.
%Invariants are key for expressing and confirming assertions about program states.
ESC/Java's emphasis on invariants helps developers express expectations precisely, allowing
potential issues to be detected early in development.
Daikon~\cite{daikon} is a well-known dynamic invariant detection tool to automatically infer likely
invariants from program executions.
Daikon takes program execution traces as input, which are typically obtained through testing.
These execution traces consist of sequences of program states and variable values observed during
the program's runtime.
%\yi{invcon should also be discussed here.}
InvCon~\cite{Liu2022Invcon} was the first tool that generates \emph{likely} invariants for smart
contracts.
With Daikon as the back-end invariant detection engine, InvCon implemented an
intermediary input transformer that coverts historic contract transactions to the compatible data
trace files accepted by Daikon.
In addition, some invariant templates are customized to support unique Solidity features, \eg,
MappingItem.
%Invariants, which are properties that remain true throughout a program's execution, are crucial for ensuring program reliability and correctness.
%
%Static invariant detection involves analyzing a program's source code without executing it, aiming to identify invariants that hold across all possible program executions. Employing techniques like abstract interpretation and theorem proving, static invariant detection is particularly useful for broader reasoning about program behavior, aiding in bug detection, optimization, and verification.
%However, due to the inherent complexity of analyzing all possible program paths statically, this approach may struggle with certain dynamic aspects of the program, potentially producing imprecise results.

There also exist other works related to invariant generation for smart contracts.
SolType~\cite{tan2022soltype} is a type checking tool for Solidity smart contracts.
It enables developers to add refinement type annotations to smart contracts, incorporating static analysis to prove that arithmetic operations are safe from integer overflows or underflows.
SolType can infer useful type annotations, but they are limited to only contract-level invariants related to arithmetic operation.
Using SolType as a verifier to learn a policy, Cider~\cite{liu2022learning} applys deep
reinforcement learning to automatically learn contract invariants.
The learned contract invariants are mainly used to guard arithmetic operations in smart contracts
to avoid integer overflows and underflows.
However, the correctness of the learned contract invariants is still not formally verified.
%\yi{have you discussed the specification repair work by Sun Jun?}

Distinguished from the aforementioned works, \tool is the first to implement a unified invariant
generation framework for Solidity contracts encompassing techniques from both dynamic detection and
static inference, where the the generated invariants are verified against the contract code.

\section{Conclusion}
\label{sec:conclusion}
We have presented \tool, a novel invariant generation framework for Solidity smart contracts where the invariants result from the integration between dynamic invariant detection and static inference.
Because implication invariants are important to capture more fine-grained program semantics of smart contracts,
\tool devises an iterative process to repeat the generation and verification of implications to overcome its combination explosion problem.
We have evaluated \tool on real-world ERC20 and ERC721 contracts and demonstrated that \tool is
able to achieve good recall to recover common specifications.
In addition, the experiments on mutation testing and vulnerable benchmark contracts have shown that
the invariant specifications generated are effective to exclude program mistakes and make contracts
secure from vulnerabilities.
%\cx{Changed from buggy to vulnerable}
%
%The invariant specification language supports only predicates over mathematical or logical
%operations on variables, in the future, we will extend this language and make \tool have the
%capability to infer other important invariants, e.g., event emission invariant and internal
%function call invariant.

% \begin{acks}
% 	This research is supported by the Singapore National Research Foundation under the National Satellite of Excellence in Mobile Systems Security and Cloud Security (NRF2018NCR-NSOE004-0001).
% \end{acks}

%\balance
\bibliographystyle{IEEEtran}
\bibliography{ref}

% \input{invtable}
%\input{sections/Appendix}
% \clearpage
% \input{invtable}
\end{document}